\documentclass[twocolumn,floatfix]{revtex4-1}
\usepackage{bm}
\usepackage{booktabs}
\usepackage{graphicx} 
\usepackage{grffile}
\usepackage{hhline}
\usepackage{pgfplots}
\usepackage{amsmath}
\usepackage{siunitx}
\usepackage{todonotes}
\presetkeys{todonotes}{inline}{}

\newcommand{\stdfigwidth}{8.9cm}

\begin{document}
\title{Interplay between particle microstructure, network topology and sample shape in magnetic gels -- A molecular dynamics simulation study}

  \author{Rudolf Weeber$^{1}$}\email{weeber@icp.uni-stuttgart.de} 
  \author{Christian Holm$^{1}$}\email{holm@icp.uni-stuttgart.de} 

\affiliation{$^1$
Institute f\"ur Computerphysik,
Universit\"{a}t Stuttgart,
Allmandring 3,
70569 Stuttgart,
Germany
}%
\date{\today}

\begin{abstract} 
Ferrogels, i.e., hydrogels loaded with magnetic nanoparticles, have the ability to deform in external magnetic fields. The precise shape of deformation and the alignment of the gel in the field, however, depend on the interplay of several factors. In this paper, we introduce a coarse-grained simulation model, which takes into account the configuration of magnetic particles in the gel, the sample shape, and aspects of the polymer network topology. We use this model to show that in gels with an isotropic microstructure, an external magnetic field reduces clustering, while this is not the case for uniaxial gels, in which the particle configuration is anisotropic due to the presence of a magnetic field during cross-linking.
For ellipsoidal gels, we find that a uniaxial microstructure additionally
can override the deformation and alignment expected as indicated the demagnetization energy.
Finally, we examine to what degree gels with different network topologies maintain the particle microstructure ``frozen in`` during the cross-linking process.
\end{abstract}


\maketitle

\section{Introduction}

Ferrogels and magnetic elastomers are soft elastic materials into which magnetic particles are embedded\cite{barsi96a,varga03a}. 
Due to the ability to control the shape\cite{szabo98a,gollwitzer08a} and elasticity\cite{mitsumata11a,odenbach16a} of such materials by means of an external magnetic field, they are candidates for a number of novel technical and biomedical applications.
Among them are actuation\cite{ramanujan06a,monz08a,zimmermann11a}, transport\cite{kondo94a,wang07d} and drug delivery\cite{hu07a,qin09b}.

The common feature of magnetic elastomers is the interplay between magnetic and elastic properties.  Details can, however, vary widely. 
For instance, the polymer matrix can either be a hydrogel or a rubber, and the elastic moduli can be tuned over two to three orders of magnitude.
Also, there is a wide choice of magnetic materials for the immersed nanoparticles, among them magnetite, maghemite, hematite, and cobalt ferrite.
It is also possible to use non-spherical particles.
Lastly, different couplings between the magnetic particles and the soft matrix can be found. 
While in some cases, the magnetic particles have some freedom to move in the gel and no direct coupling exists between the orientation of the magnetic moment and the matrix, in other cases\cite{messing11a,roeder14a}, a rotation of the magnetic moment will directly result in a strain in the polymer matrix.
Considering the large flexibility in making the materials, it is not surprising that some of the observations made in experiments are contradictory.
A solid understanding of the microscopic properties of the materials is needed to explain experimental observations and, in the future, make predictions about materials not yet synthesized.

In recent years, many aspects of ferrogels have been studied. For instance, elastic properties\cite{mitsumata11a,wood11a,annunziata13a,cremer15a}, the deformation in a field\cite{raikher03a,gollwitzer08a,weeber12a,weeber15a,weeber15c}, the influence of the particle configuration in the gel\cite{collin03a,stolbov11a,ivaneyko12a,allahyarov15a}, and magnetic particles used as cross-linkers\cite{messing11a,ilg13a,roeder14a,roeder15a}.

Depending on the question under consideration, the modelling of magnetic elastomers can be carried out on different length scales. On the most detailed level, the polymers are modeled as bead-spring-systems\cite{dudek07a,weeber12a,weeber15a,weeber15c,pessot15a}, which is useful for detailed studies of, e.g., the polymer-nanoparticle coupling, but is limited to very small systems. On the next coarser level, it is possible to model the polymers as spring networks\cite{huang16a,cremer15a}. Still larger samples, up to the macroscopic scale, can be studied by treating the polymer matrix as an elastic continuum\cite{wood11a,raikher03a,stolbov11a,zubarev12a,brand14a},
which is helpful, e.g., for studying the influence of the sample shape and for developing models which can be treated analytically.

Four factors can determine the response of a ferrogel in a magnetic field: the sample shape\cite{raikher05a,zubarev12a}, the microscopic configuration of the nanoparticles\cite{zubarev12a,pessot14a},  the coupling between the polymers and the nanoparticles\cite{weeber12a,weeber15a}, and the topology of the polymer network\cite{weeber15c}.
It is the aim of this investigation to include three of these contributions into a single model, namely the sample shape, the particle microstructure and the network topology.
We place emphasis on a realistic magnetic particle configuration, which we obtain from simulations of a ferrofluid.
By applying a magnetic field during these simulations, we can also generate microstructures for uniaxial ferrogels\cite{collin03a,menzel14a,brand14a}, i.e., those for which the particle microstructure exhibits a preferred direction.
We include up to 10\,000 magnetic particles, so that the ``macroscopic'' sample shape effects can be captured. In our simulations, we employ volume fractions and magnetic moments for the nanoparticles which should be experimentally viable.
In terms of the topology of the polymer network, we compare three scenarios. In two of these, the probability for two magnetic particles to be linked by a polymer is highest for a particle-particle distance of zero.  We associate this with a situation, in which the magnetic particles attach to the size of polymer chains. One of the two cases makes use of a narrow, the other of a wide distribution function.
The third case pertains to gels in which the magnetic particles are attached to the end of the polymer chains. Here, the highest probability for two particles to become connected is realized at a particle-particle distance larger than zero. 
The remainder of the paper is structured as follows: in Sec.\,\ref{sec:model}, we introduce the simulation model and procedure. Then, the influence of the microstructure (Sec.\,\ref{sec:microstructure}) and the sample shape (Sec.\,\ref{sec:shape}) will be discussed. The findings are supplemented by an analysis of the gels' magnetic response in Sec.\,\ref{sec:mag}. Lastly, we study the influence of the polymer network topology in Sec.\,\ref{sec:topology} and put it all together in a summary.

\section{Model}
\label{sec:model}

The parameters for the gels considered here are chosen in a range, which should be experimentally viable. For the magnetic particles, we use a volume fraction of 5\% and a dipolar interaction parameter of $\lambda =4$. This parameter compares the maximum absolute value of the dipole-dipole interaction between two touching magnetic particles to the thermal energy. It is given by
\begin{equation}
\label{eqn:lambda}
\lambda =\frac{\mu_0 m^2}{4 \pi \sigma^3 k_B T},
\end{equation}
where $\mu_0$ is the vacuum permittivity, $m$ the particles magnetic moment $\sigma$ the particle diameter and $k_B T$ the thermal energy.
We use magnetic fields with a Langevin parameter of $\alpha=15$, where
\begin{equation}
\alpha =\frac{\mu_0 m H}{k_B T}
\end{equation}
compares the Zeeman energy to the thermal energy. 
Here $\mu_0$ denotes the vacuum permittivity, $m$ the magnetic moment of a particle and $H$ the magnetic field.

For studying the influence of the microstructure on the gels' response to a magnetic field, the magnetic particles have to be explicitly modelled in the simulation. Simultaneously, the sample has to be large enough for us to be able to describe its shape and the corresponding influence of the demagnetization field (the magnetic field created by the magnetization of the sample itself). To satisfy both conditions, we use approximately 10\,000 particles in the spherical case. The polymers are approximated by harmonic springs, because explicitly modelling them in a bead-spring framework as in Refs.\,\cite{weeber12a,weeber15a,pessot15a,weeber15c} would result in too large of a computational effort. 
At the given density the diameter of the spherical gel sample is approximately 60 times diameters of a magnetic particle.

We carry out the modelling in four steps, namely obtaining the microstructure of a ferrofluid, cutting the sample shape, cross-linking and finally obtaining observables for the gel.
Here, we give an overview of the steps. The details are found in Appendix \ref{sec:sim}. 

\begin{figure}
\includegraphics[angle=270,width=\stdfigwidth]{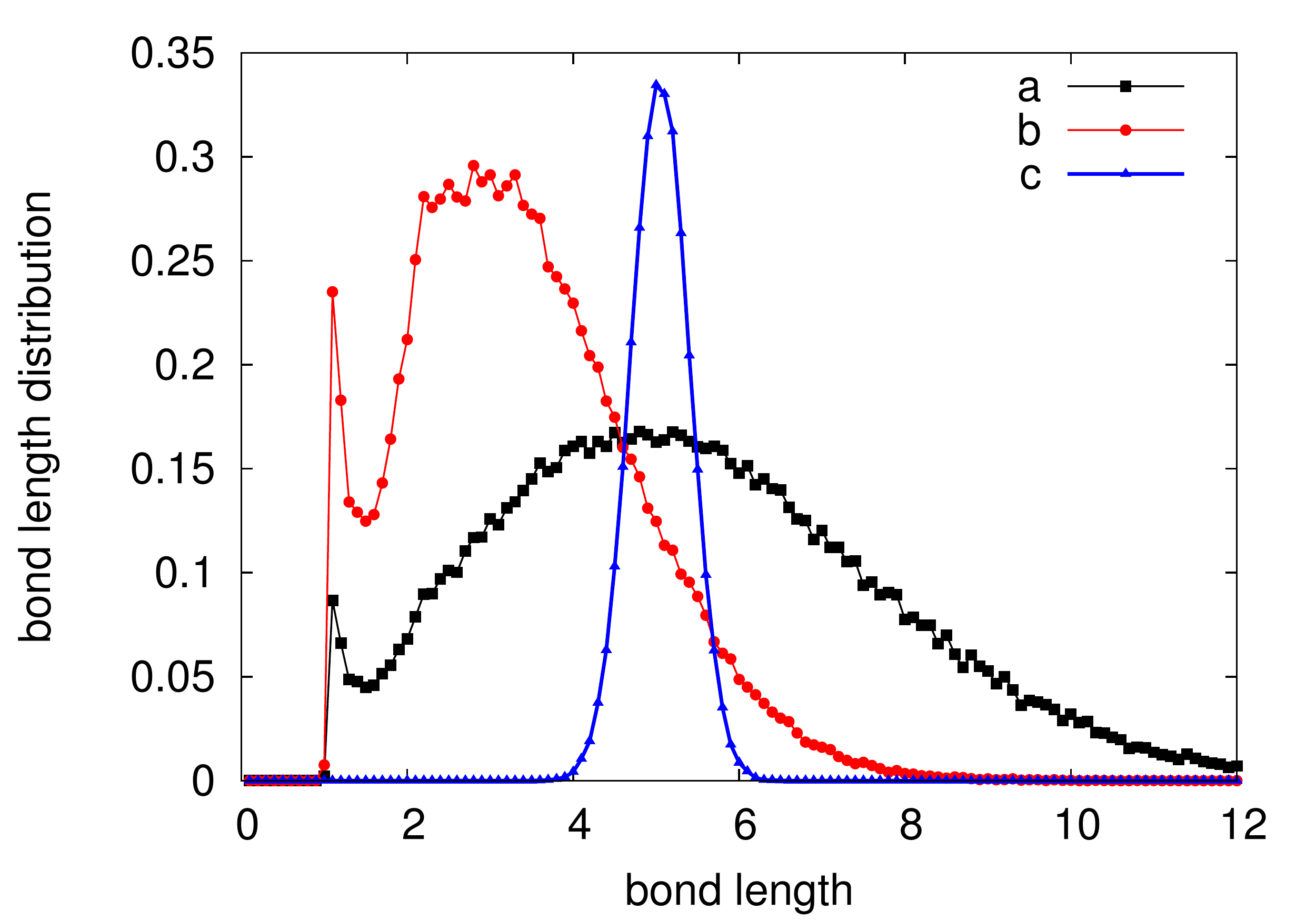}
\caption{\label{fig:bond-length-dist}
Distribution of the bond length (average end-to-end distance of the polymers, $l_b(r)$) for the three cases $a$ through $c$ (Eqns.\,\ref{eqn:bond-prob}-\ref{eqn:bond-stiffness} and Tbl.\,\ref{tbl:bond-parameters}).
The peak slightly above one particle diameter arises from the corresponding peak in the pair-correlation function of a ferrofluid. 
}
\end{figure}

The procedure is as follows:
First, a realistic distribution of the magnetic particles at the time of cross-linking is obtained from a molecular dynamics simulation of a standard ferrofluid. This simulation is carried out at the desired density with periodic boundary conditions applied. If a uniaxial gel is to be constructed, an external field is applied, resulting in a non-isotropic distribution of magnetic particles.
The simulations are very similar to previously performed simulations, e.g. in Refs.\,\cite{ilg05a,cerda08a,jager11a,weeber13a}.
Once a configuration of magnetic particles is obtained, in the second step, the desired sample shape is cut from the ferrofluid. In this paper, we consider spheres, as well as oblate and prolate ellipsoids of revolution (oblate: one short and two equal long axes, prolate: one long and two equal short axes).
In the third step, the gel is cross-linked by adding polymers represented as soft harmonic ``springs`` between pairs of magnetic particles. 
The harmonic potential is applied to the centers of the magnetic particles. Hence, the bond does not constrain the particles'  rotation.
Two choices have to be made, namely the probability that two magnetic particles are attached to the same polymer and the elasticity of the polymer.
In our model, both of these are controlled by distance-dependent functions.
For the bonding probability, we use
\begin{equation}
\label{eqn:bond-prob}
p_b(r) = \min\left(1,c_1 e^{-(r-r^{\ast})^2/c_2^2}\right).
\end{equation}
Setting $r^{\ast}=0$ will result in a peak of the probability at $r=0$, which we associate with particles binding at the ``side'' of polymer chains, as mentioned above. The parameter $c_2$ controls the width of the distribution and can be linked qualitatively with the polymers' end-to-end length distribution.
The actual bond length distribution can be estimated by multiplying the number of neighboring particles at a certain distance with the bond probability function, i.e.,
\begin{equation}
\label{eqn:bond-length-dist}
l_b(r) =4\pi r^2 \phi g(r) p_b(r),
\end{equation}
where $\phi$ is the number density and $g(r)$ is the pair-correlation function of the cross-linked gel. 
Bond length distributions for the three topologies we study are shown in Fig.\,\ref{fig:bond-length-dist}. They have been obtained from the actual systems being simulated.

The stiffness of the bond is chosen inverse proportional to its lengths, as there is more flexibility in a long polymer chain than in a short one.
The spring constant of the harmonic bond is given by
\begin{equation}
\label{eqn:bond-stiffness}
k(r_0) =k_0/r_0,
\end{equation}
where $r_0$ denotes the equilibrium bond length, which is identical to the inter-particle distance at the time of cross-linking.
The parameters used for the three network topologies we compare are summarized in Table \ref{tbl:bond-parameters}. 
The spring constants are chosen in such a way that the magnetic particles retain some freedom of motion.
As all bonds are added in such a way that the particle configuration at the time of cross-linking minimizes the potential energy, there is no significant deformation during the following simulations, unless an external field is applied.

\begin{figure*}
\includegraphics[width=\stdfigwidth]{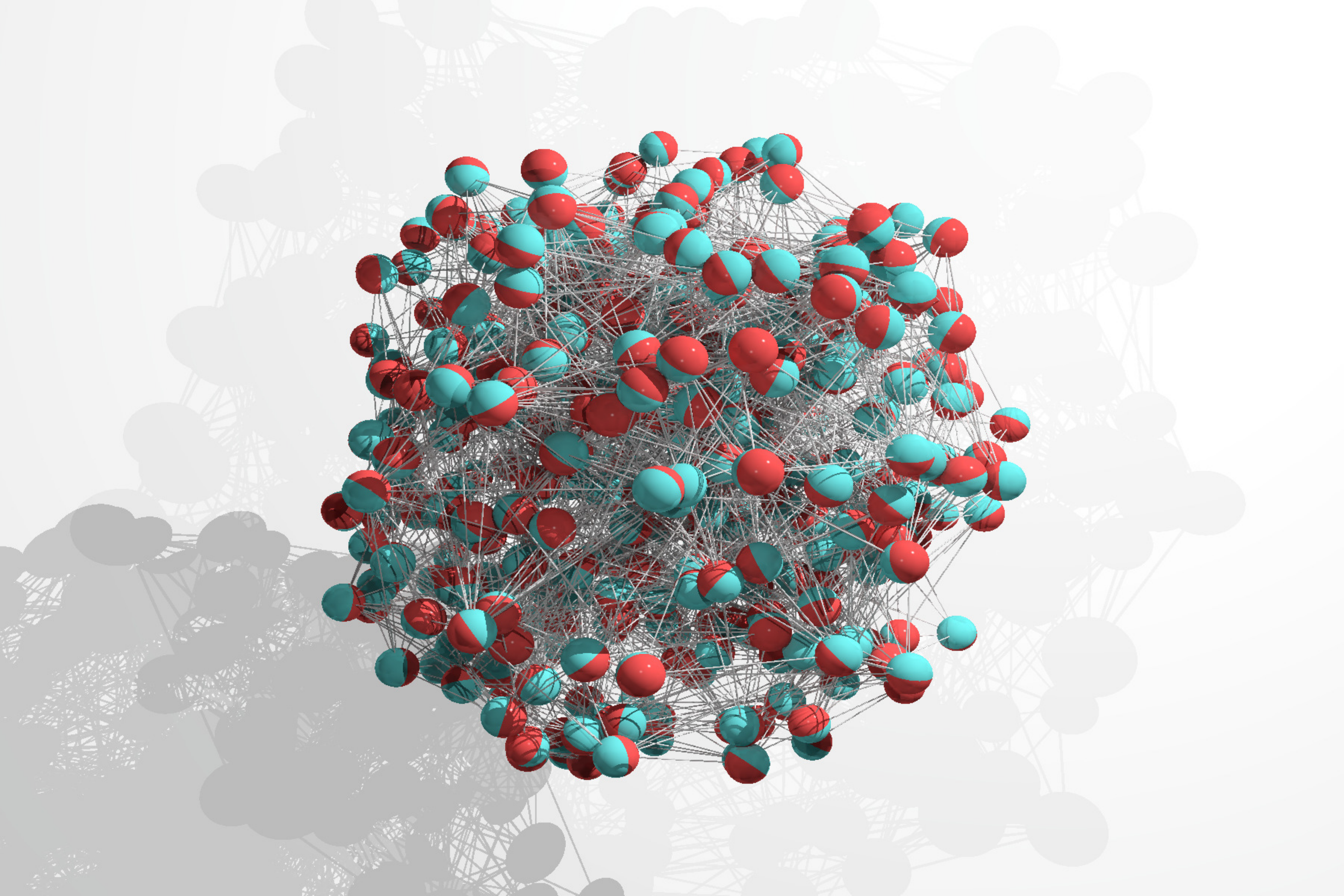}
\includegraphics[width=\stdfigwidth]{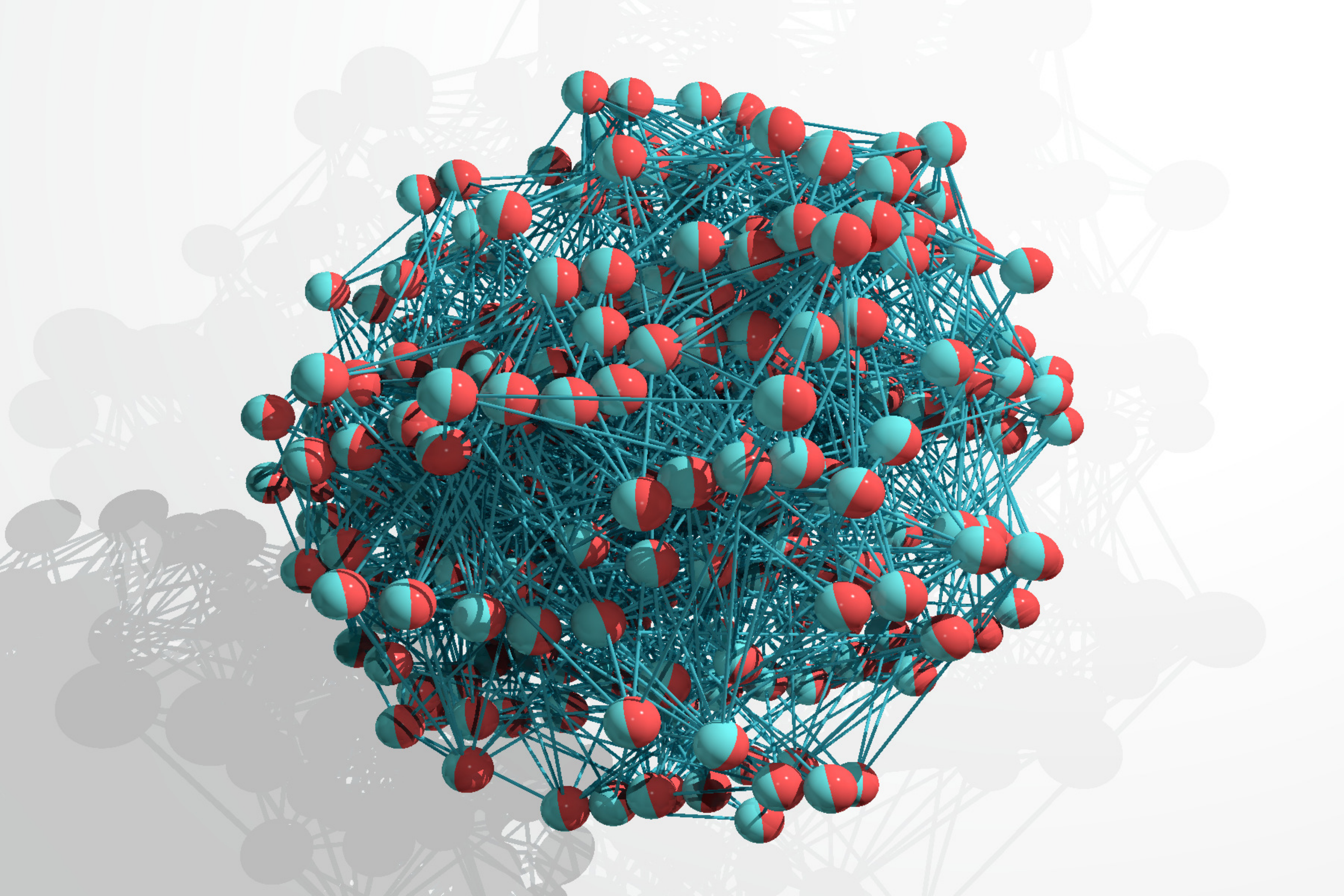}
\caption{\label{fig:snapshot}
Snapshots of small gel samples cross-linked without an external magnetic field (left) and in a field of $\alpha=15$ (right). 
The gels shown here contain approximately 2\% of the particles used in the simulations throughout the rest of the paper. The snapshots are taken immediately after cross-linking.
When there is no external field during cross-linking, the gel's microstructure is isotropic. Applying a field during cross-linking results in a uniaxial microstructure, as chains-like clusters of magnetic particles are ``frozen in``.
}
\end{figure*}

In the final step, the resulting gel is simulated using a molecular dynamics and Monte Carlo hybrid method, with open boundary conditions to measure observables. This can be done at an external magnetic field different from the one applied during cross-linking.
Snapshots of gel samples with isotropic and uniaxial microstructures are compared in Fig.\,\ref{fig:snapshot}.
To improve visibility, small systems are shown, which contain only approximately 2\% of the particles used in the full simulations discussed throughout the paper.
A technical description of the individual simulation steps is provided in Appendix \ref{sec:sim}.

\begin{table}
\centering
\begin{tabular}{%
  S[table-format=1]||%
  S[table-format=1]|%
  S[table-format=1.3]|%
  S[table-format=1.1]|%
  S[table-format=1.2]}
{case} & {$r^{\ast}$} & {$c_1$} & {$c_2$} & {$k_0$} \\
\hhline{=#=|=|=|=}
a & 0 & 0.4  & 5 & 5 \\
\hline
b & 0 & 1.732 & 3 & 9.07 \\
\hline
c & 5 & 0.995 & 0.5 & 5 \\
\end{tabular}
\caption{
\label{tbl:bond-parameters}
Parameters for Eqns.\,\ref{eqn:bond-prob} and \ref{eqn:bond-stiffness} controlling the bonds between the magnetic particles, and thus, the resulting network topology. 
Three cases are considered. For cases a and b, the probability function (Eqn.\,\ref{eqn:bond-prob}) peaks at zero, which we associate with magnetic particles attaching to the side of polymers. Case c pertains to a situation where the particles connect to the end of the chains. Hence the probability function peaks at a value larger than zero.
For the three cases, the number of bonds per particle and the elasticity are matched. 
}
\end{table}
\begin{figure}
\includegraphics[angle=270,width=\stdfigwidth]{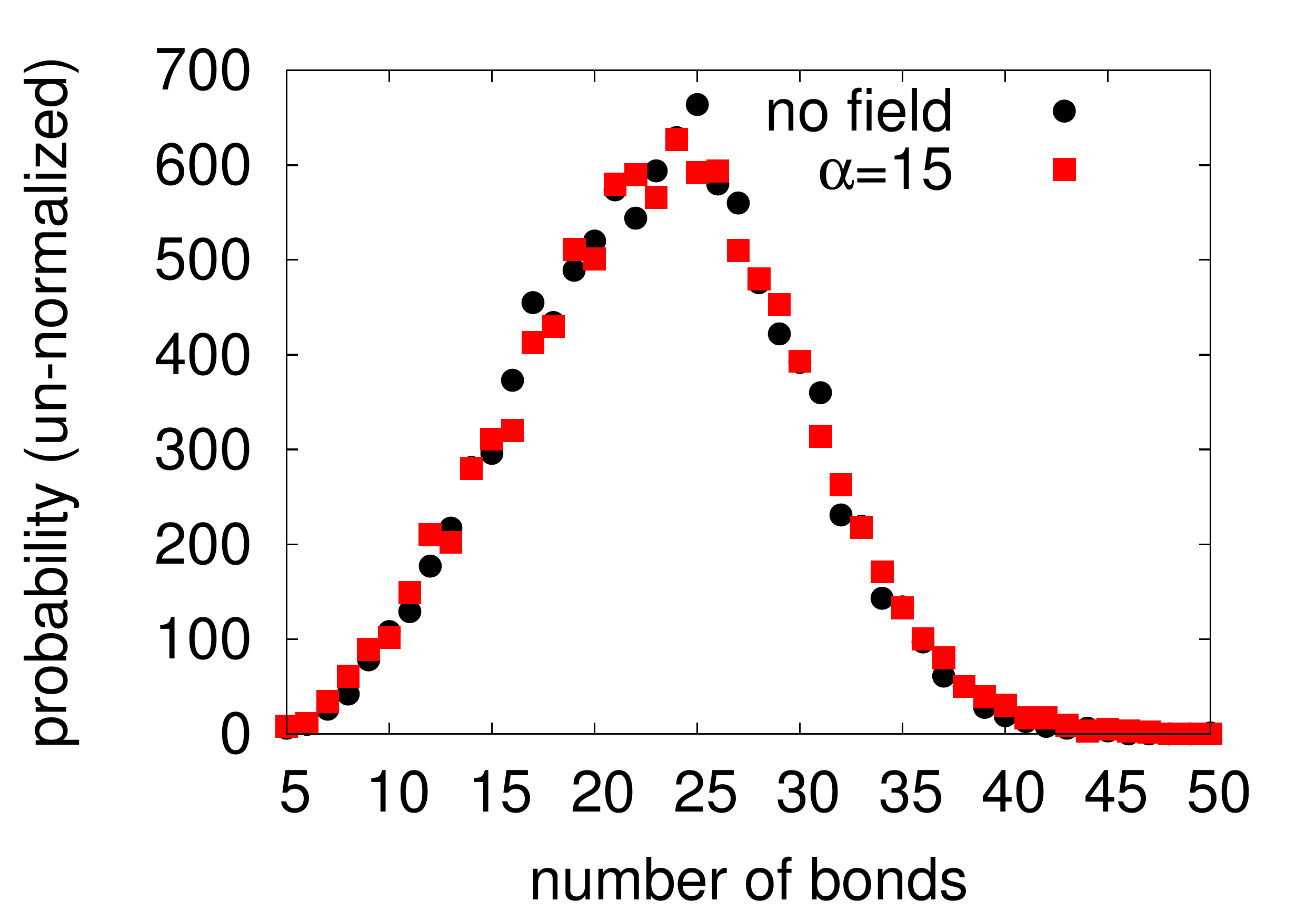}
\caption{\label{fig:bond-dist}
Distribution of the number of polymers connected to a magnetic particle.
The prefactors ($c_1$ in Eqn.\,\ref{eqn:bond-prob}) in table \ref{tbl:bond-parameters} are chosen such that the average number of bonds per particle is approximately the same for all three topologies considered. Additionally, the samples' elasticities have been matched to each other.
}
\end{figure}

\section{Influence of the microstructure on the deformation}
\label{sec:microstructure}

\begin{figure*}
\includegraphics[angle=270,width=\stdfigwidth]{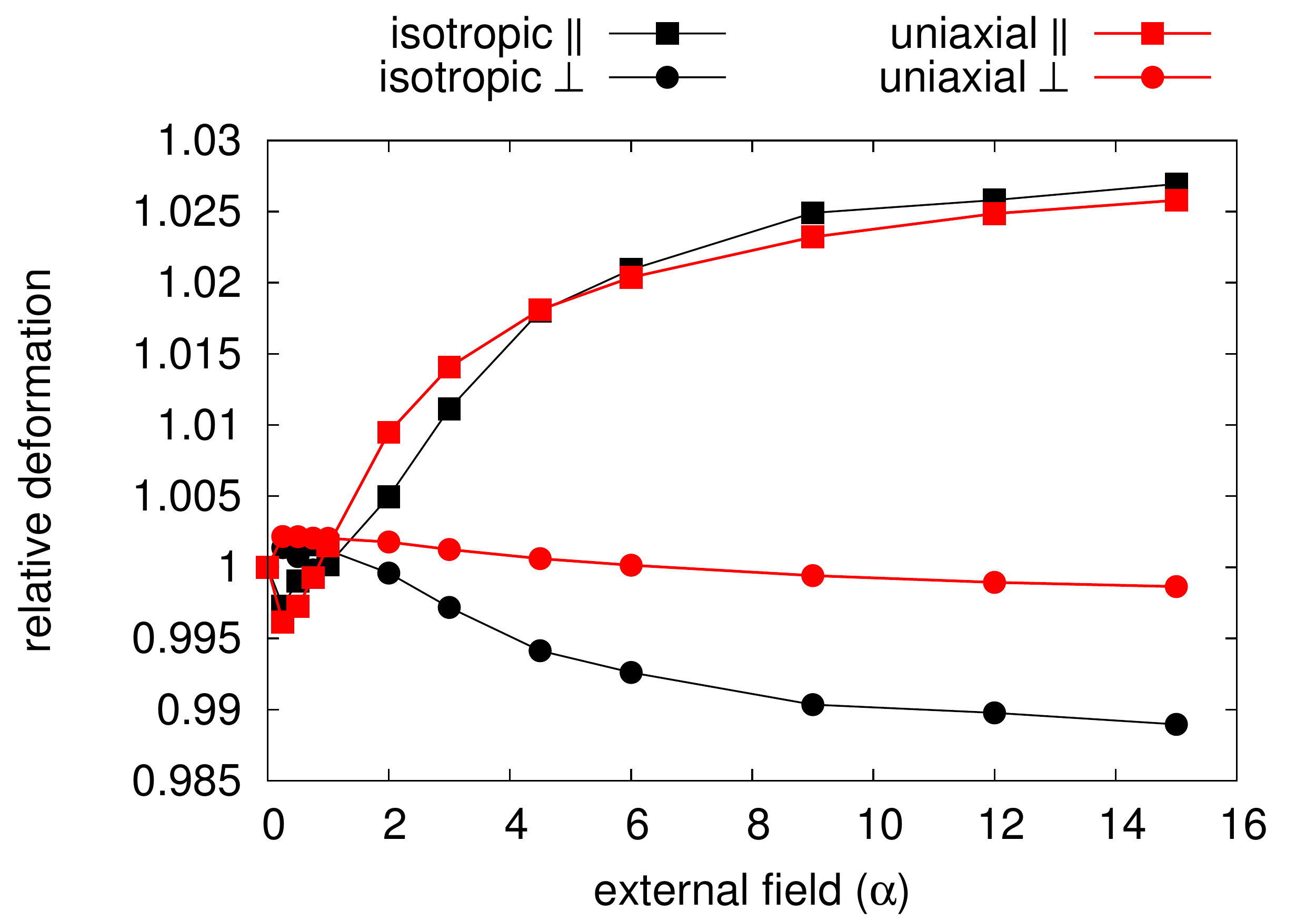}
\includegraphics[angle=270,width=\stdfigwidth]{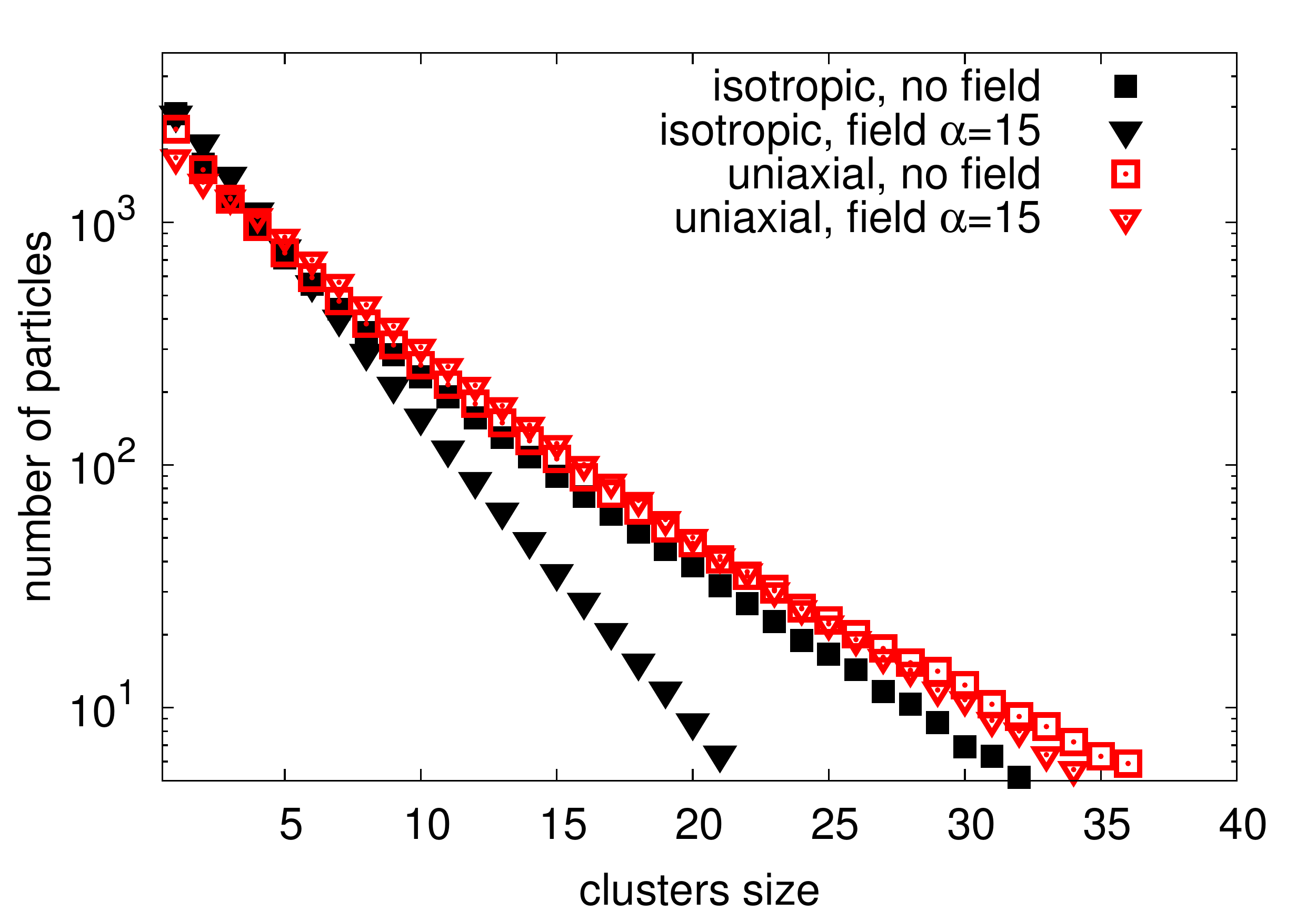}
\caption{\label{fig:spherical}
{\bf Left:} Relative deformation of spherical gel samples with isotropic and uniaxial microstructures versus the applied magnetic field. The $\parallel$ and $\perp$ signs indicate the orientation relative to the field.
For both microstructures, the gel elongates parallel to the applied field. A shrinkage in the perpendicular direction is observed for the isotropic case, whereas the uniaxial gel does not deform significantly in the perpendicular direction.\newline
{\bf Right:} Total number of particles which are part of clusters of a given cluster size (Eqn.\,\ref{eqn:particles-in-clusters}) for isotropic and uniaxial gels in the field free case and for an applied field of $\alpha=15$. In the isotropic case, applying an external field reduces clustering, as clusters not oriented parallel to the field are energetically unfavorable. In the uniaxial case, much more large clusters are found even in the field free case, and the clustering is not changed significantly by applying an external field.
}
\end{figure*}

Let us first consider the influence of the microstructure on the gels' deformation in an external field. 
On the left hand side of Fig.\,\ref{fig:spherical}, the relative deformation for directions parallel and perpendicular to the applied field is shown for a spherical gel sample.
The shape is determined by comparing the gels' moments of inertia around the Cartesian axes to those of an ellipsoid (See Appendix \ref{sec:det-shape} for details).
It can be seen that both, for an isotropic and a uniaxial microstructure, the gel expands in the direction parallel to the applied field. 
On the macroscopic level, this is associated with a reduction in the sample's demagnetization energy: for homogeneously magnetized samples, it is more favorable to elongate along the direction of the magnetization and shrink in the perpendicular direction.
On the microscopic level, it can be attributed to chains of magnetic particles bending into the field direction. Also, a shape elongated parallel to the external field, and therefore also parallel to the dipole moments in the sample, increases the number of pairs of particles, for which the dipole-dipole interaction is favorable.

In the directions perpendicular to the field, the behaviour differs depending on the gels' microstructure. While there is a significant shrinkage for gels with isotropic microstructure, no notable shrinkage is observed in the uniaxial case.
There might be two reasons for this. First, there are more and longer chains in the gels based on a uniaxial microstructure. These chains repel each other in the lateral direction, which may result in an increased stiffness of the gel.
Secondly, in the uniaxial case, the chains all tend to be aligned along one single direction, namely the field direction during cross-linking. When a field is applied, the entire sample rotates to re-align this direction to the applied field. 
In the isotropic gel, on the other hand, the chains are aligned along various directions. When a field is applied, some of these chains will bend into the field direction, which may result in a shrinkage in the perpendicular direction.
Further insight into the microstructure of the gel can be derived from a cluster analysis.
The analysis is performed in analogy to Refs.\,\cite{klinkigt13a,weeber13a}. Two magnetic particles are taken to be part of the same cluster if their distance is smaller than $1.3 \sigma$ and the dipole-dipole interaction between them is less or equal to zero.
The plot on the right-hand side of Fig.\,\ref{fig:spherical} shows the number of particles $N(s)$ which are part of clusters of size $s$. 
Denoting the number of clusters of size $s$ by $n(s)$, we have
\begin{equation}
\label{eqn:particles-in-clusters}
N(s) =s \, n(s).
\end{equation}
The cluster size distribution is averaged over 4800 individual snapshots.
When the gel has a uniaxial microstructure (open symbols in Fig.\,\ref{fig:spherical}), more particles are part of large clusters than in the isotropic case.
Also, as there are many long chains aligned to a single direction already in the field-free case, the constitution of the system does not change significantly, when a field is applied.
In a gel with an isotropic microstructure on the other hand, chains are aligned in various directions. When a field is applied, some of these will bend into the field direction to minimize the Zeeman energy, but for others the cost in elastic energy will be too large. Hence, these chains will break apart into smaller pieces, which can individually align with the field, thereby reducing the clustering (full symbols in Fig.\,\ref{fig:spherical}).
Thus, for gels with an isotropic microstructure, applying an external field can dampen clustering.
This is the opposite of the behaviour observed in standard ferrofluids, where applying a field leads to extended clustering.

\begin{figure*}
\begin{minipage}[t]{0.48\linewidth}
\begin{center}
Uniaxial gel, cross-linked in a field of $\alpha=15$\\
\includegraphics[width=\linewidth]{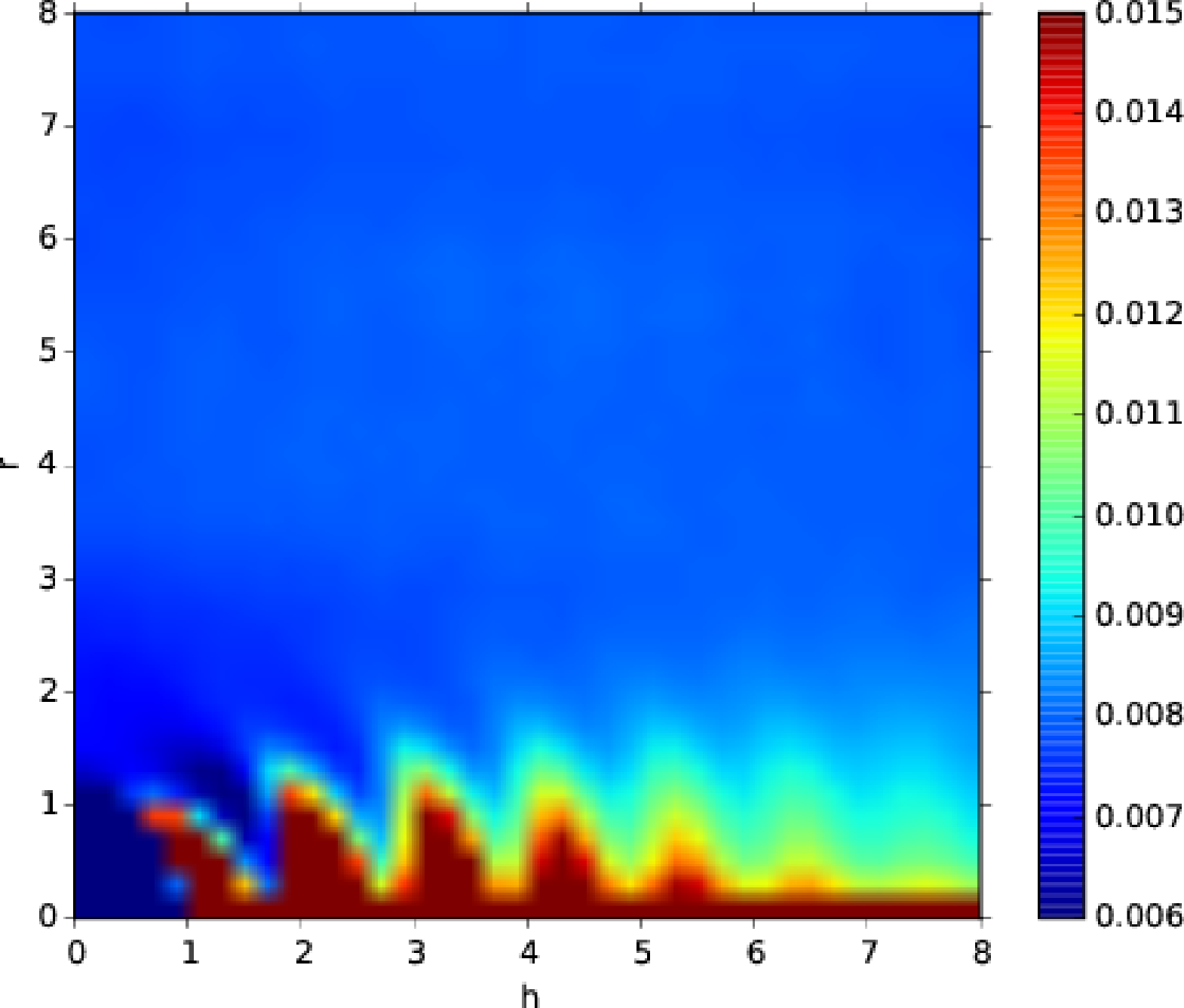}
\end{center}
\end{minipage}
\begin{minipage}[t]{0.48\linewidth}
\begin{center}
Isotropic gel, no field during cross-linking\\
\includegraphics[width=\linewidth]{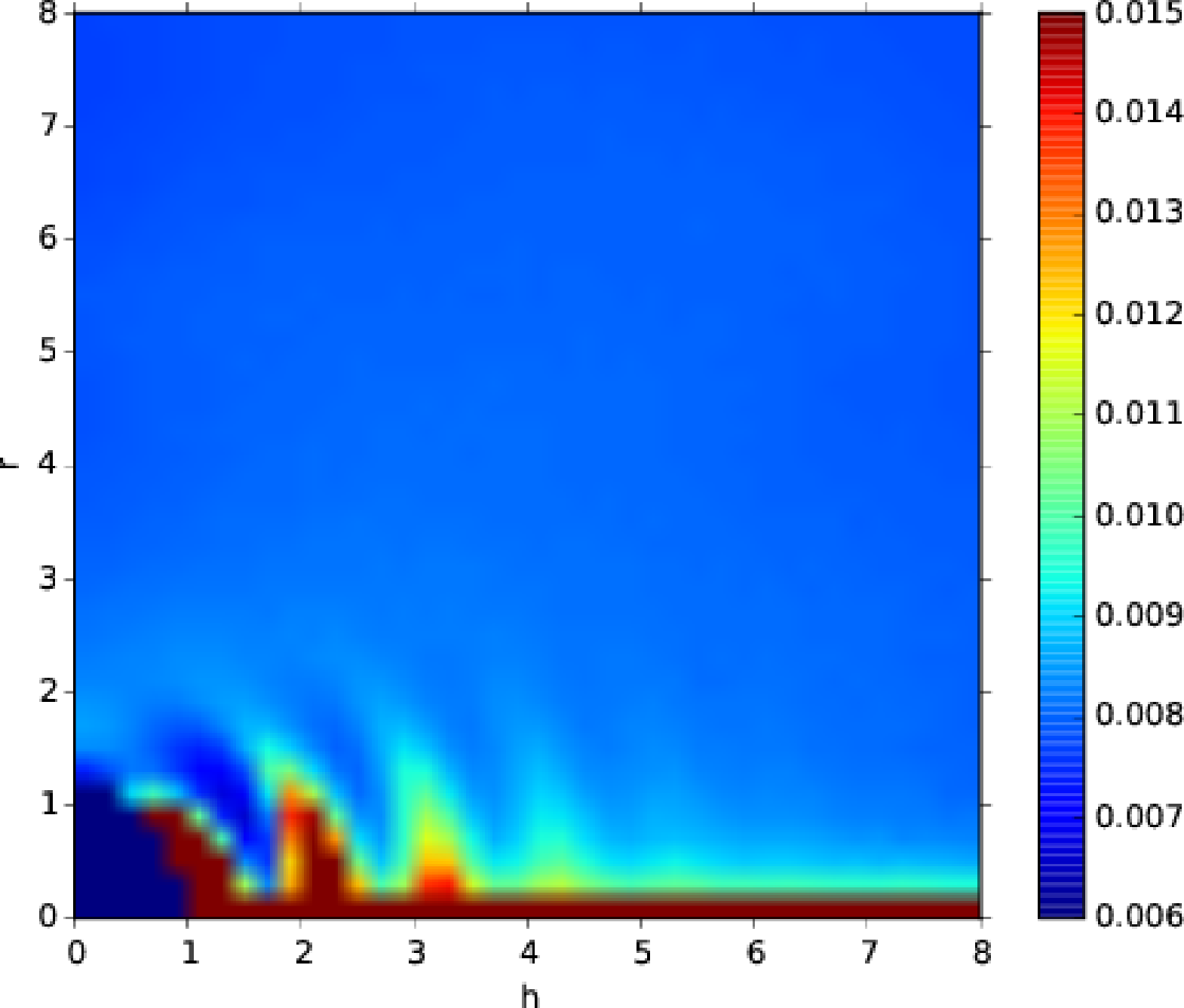}
\end{center}
\end{minipage}
\caption{\label{fig:correlation}
Pair correlation $g(h,r)$ for uniaxial and isotropic gels in a magnetic field of $\alpha=15$.
The symbol $h$ denotes the distance in field direction and $r$ the radial distance.
In the uniaxial case, strong chain formation parallel to the field causes peaks in the correlation with a spacing of approximately one particle diameter, whereas there is a depletion zone in the radial direction. For an isotropic gels, the peaks are less pronounced due to lower chaining. Also, the depletion zone is not present, as a significant number of chains are not aligned parallel to the field.
}
\end{figure*}
Further insight into the microstructure of the magnetic particles can be obtained from pair-correlation functions. When considering the case of an external field being applied, it is helpful to use a two-dimensional correlation function $g(r,h)$, which denotes the average density at a given distance $h$ in field direction and $r$ in the perpendicular direction from a particle. The resulting data for uniaxial and isotropic gels placed into a field of $\alpha=15$ is shown in Fig.\,\ref{fig:correlation}. Due to the rotational symmetry around the field axis and the mirror symmetry with respect to the $h=0$ line, only one quadrant is shown. Blue denotes a low density, and red a high one. Values smaller and larger than the lower and upper value on the scale bar are drawn in blue and red, respectively.
As can be seen in the figure, in steps of approximately a particle diameter in field direction, the density around a particle shows peaks. This is a signature of the formation of chains aligned parallel to the external field. It can also be seen that to the side of a particle, for $h<\sigma$ and $\sigma <r \lesssim 2\sigma$,
there is a depletion zone forming because it is energetically unfavorable for two magnetic particles in the side-by-side configuration to have their dipole moments aligned in parallel. 
Comparing the uniaxial (left) and isotropic (right) case, we can see that there is much less clustering in the isotropic case, which is in agreement with the results shown on the right hand side of Fig.\,\ref{fig:spherical}. Also, in the isotropic case, the depletion next to the clusters is less pronounced. This may be an indication that some chains could not align to the field direction, as the cost in elastic energy is too high. The results in Fig.\,\ref{fig:spherical} suggest that some of these chains break apart.

\section{Sample shape}
\label{sec:shape}

Due to the demagnetization field, the energy of the system depends on the shape of the sample as a whole. In many cases, a homogeneously magnetized elastic medium tends to elongate in the field direction and shrink in the perpendicular direction\cite{raikher03a,gollwitzer08a}. 
If, however, like in the case of a ferrogel, the system is structured on a microscopic level, there are situations when this is no longer the case.
In this section, we will therefore study the interplay between the microstructure and the sample shape.
As in the previous section, we consider uniaxial and isotropic microstructures. We chose two sample shapes, namely prolate (one long and two short axes) and an oblate (one short and two long axes) ellipsoids of revolution. 
For both shapes, we chose the long axes to be equal to the diameter of the spherical system studied so far, and the short axes such that the volume of the ellipsoid equals half that of the spherical system.

\begin{figure}
\includegraphics[width=0.26\linewidth]{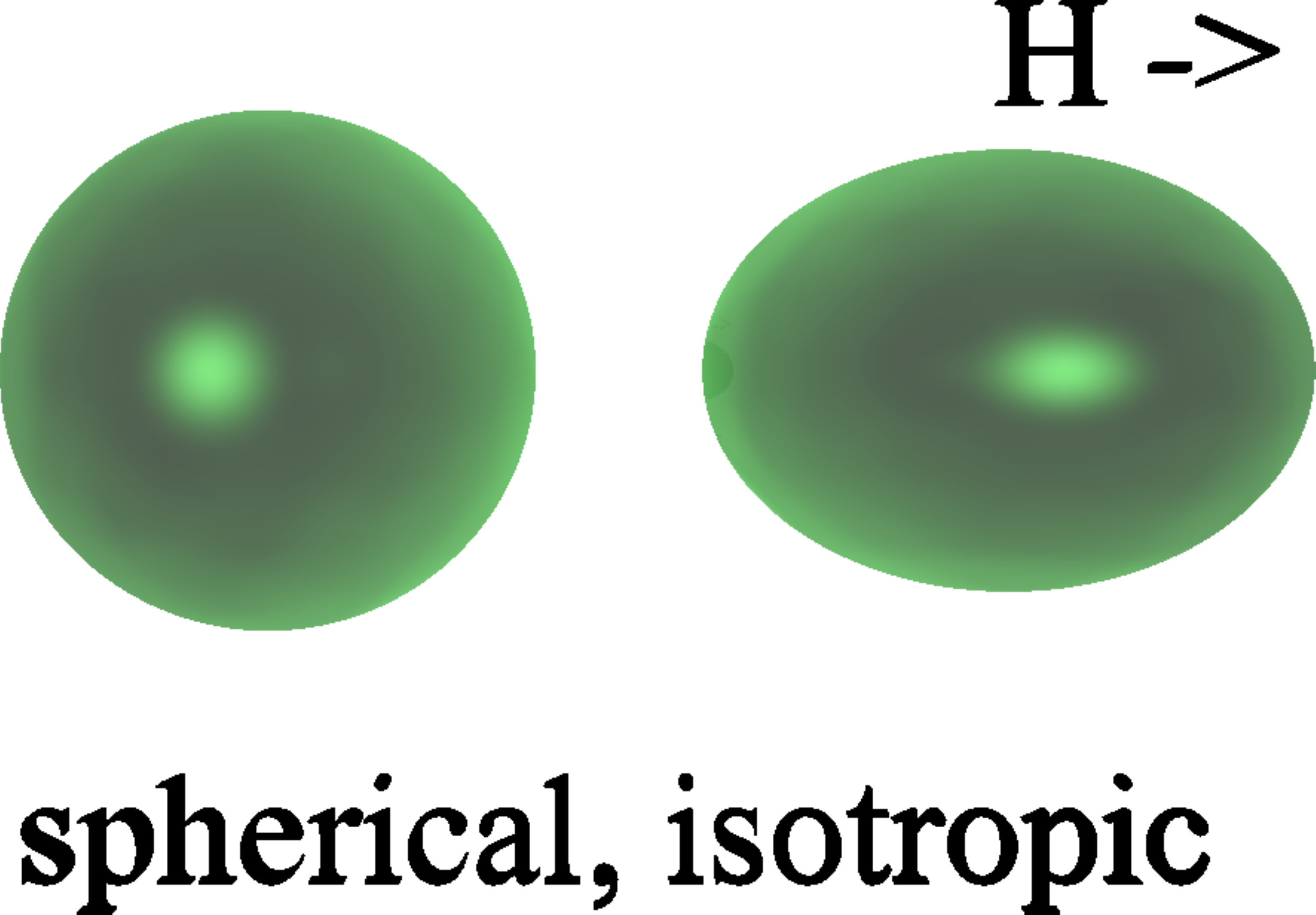}
\hspace{1em}
\includegraphics[width=0.26\linewidth]{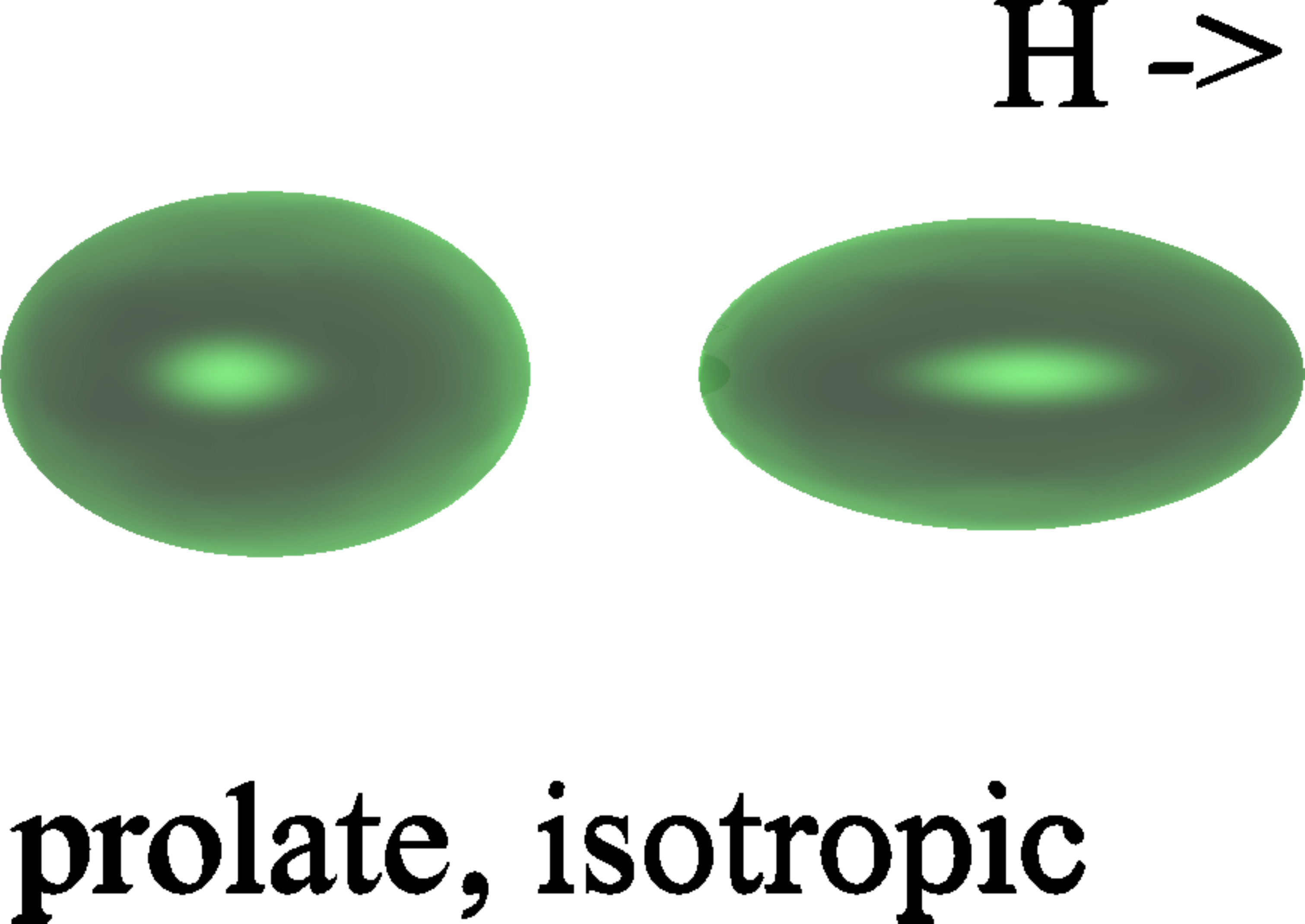}
\hspace{1em}
\includegraphics[width=0.26\linewidth]{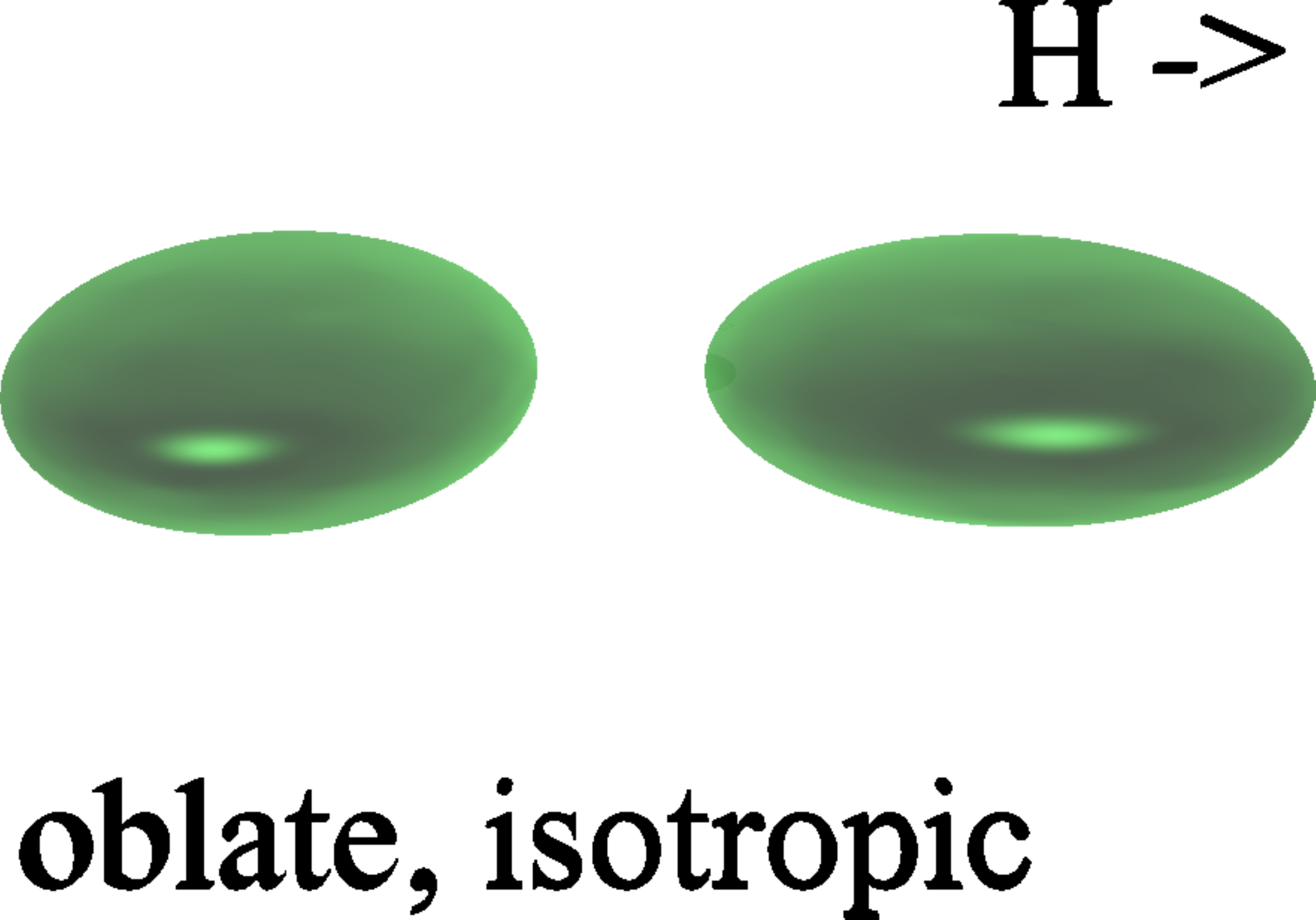}\\
\includegraphics[width=0.26\linewidth]{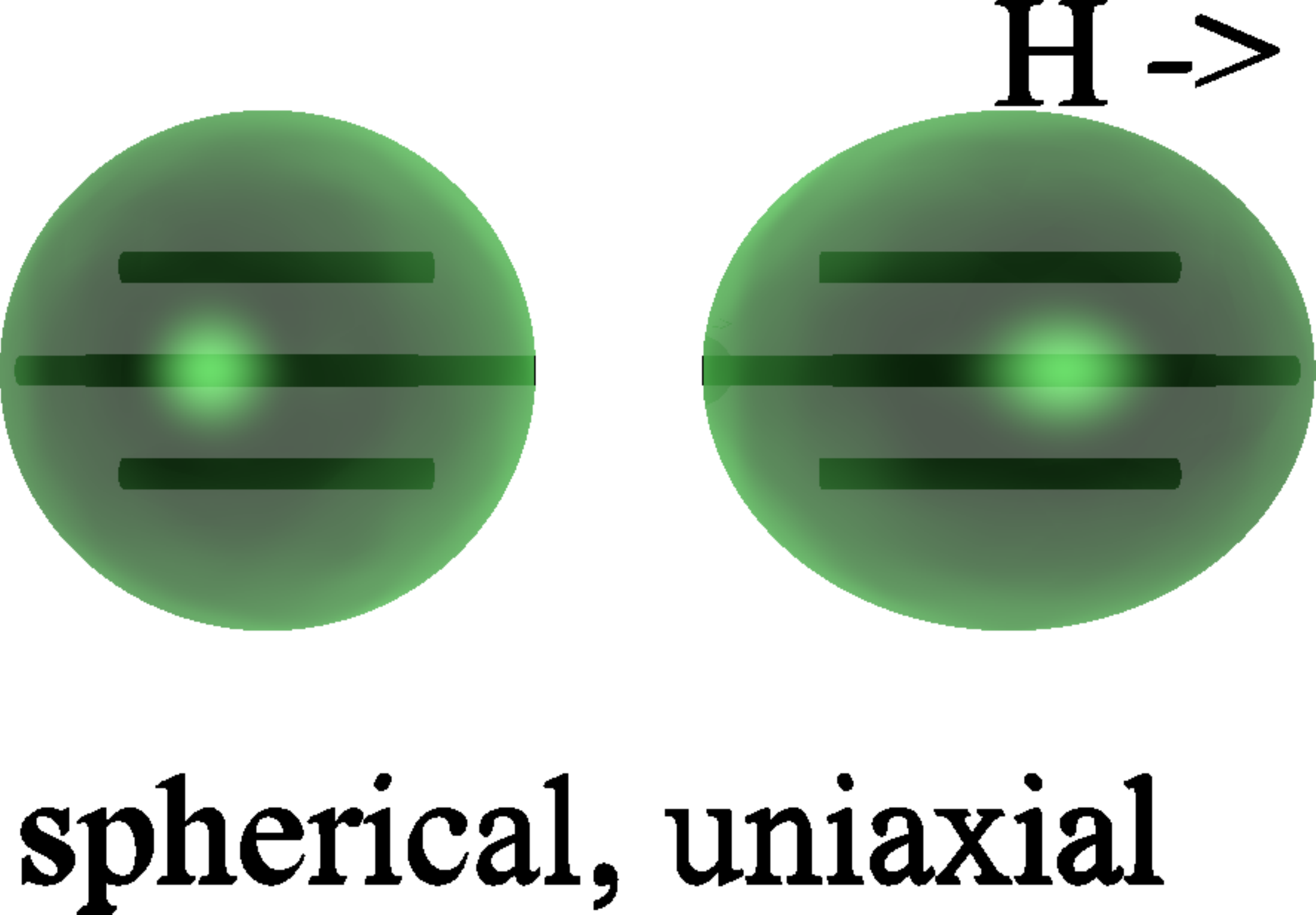}
\hspace{1em}
\includegraphics[width=0.26\linewidth]{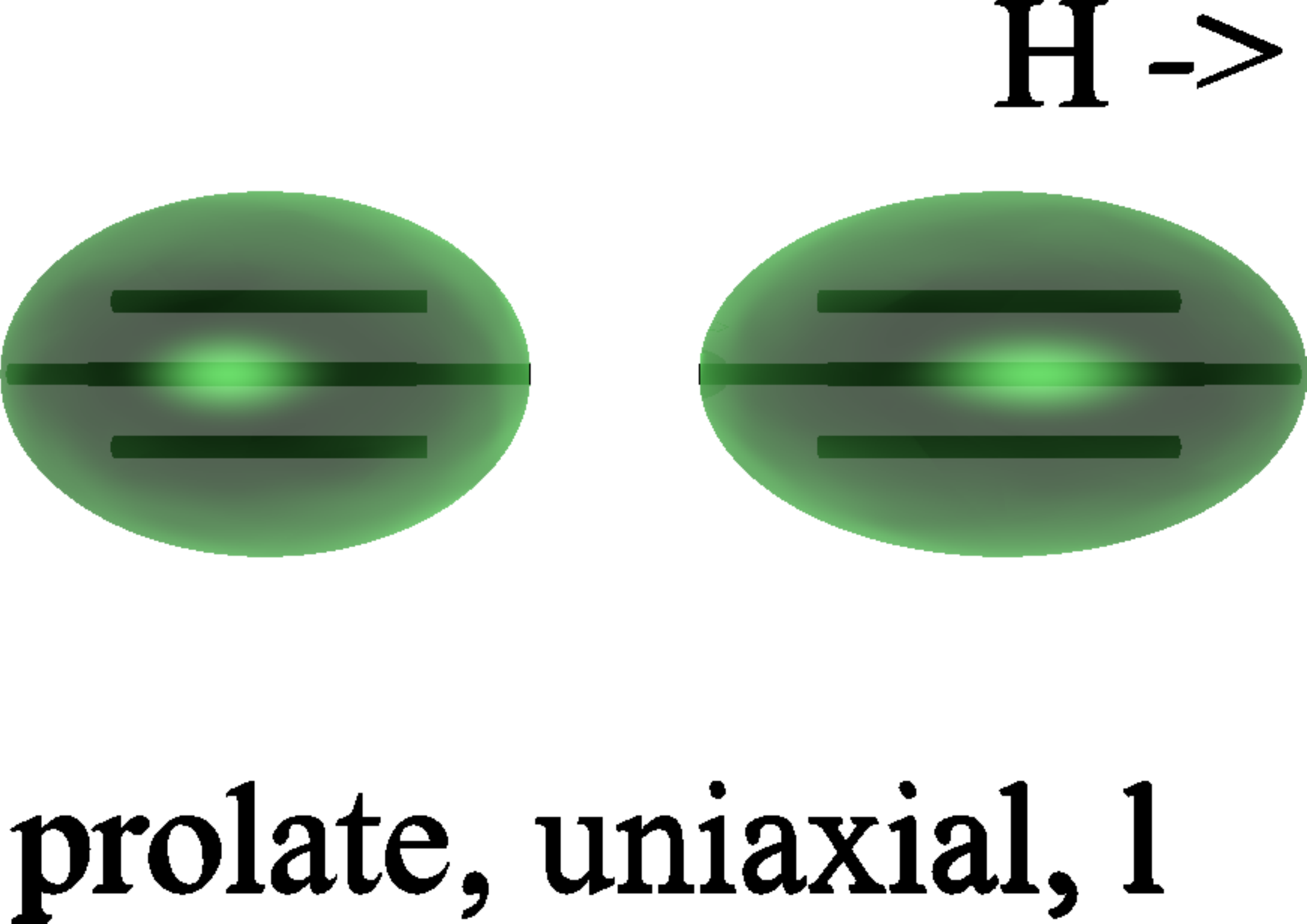}
\hspace{1em}
\includegraphics[width=0.26\linewidth]{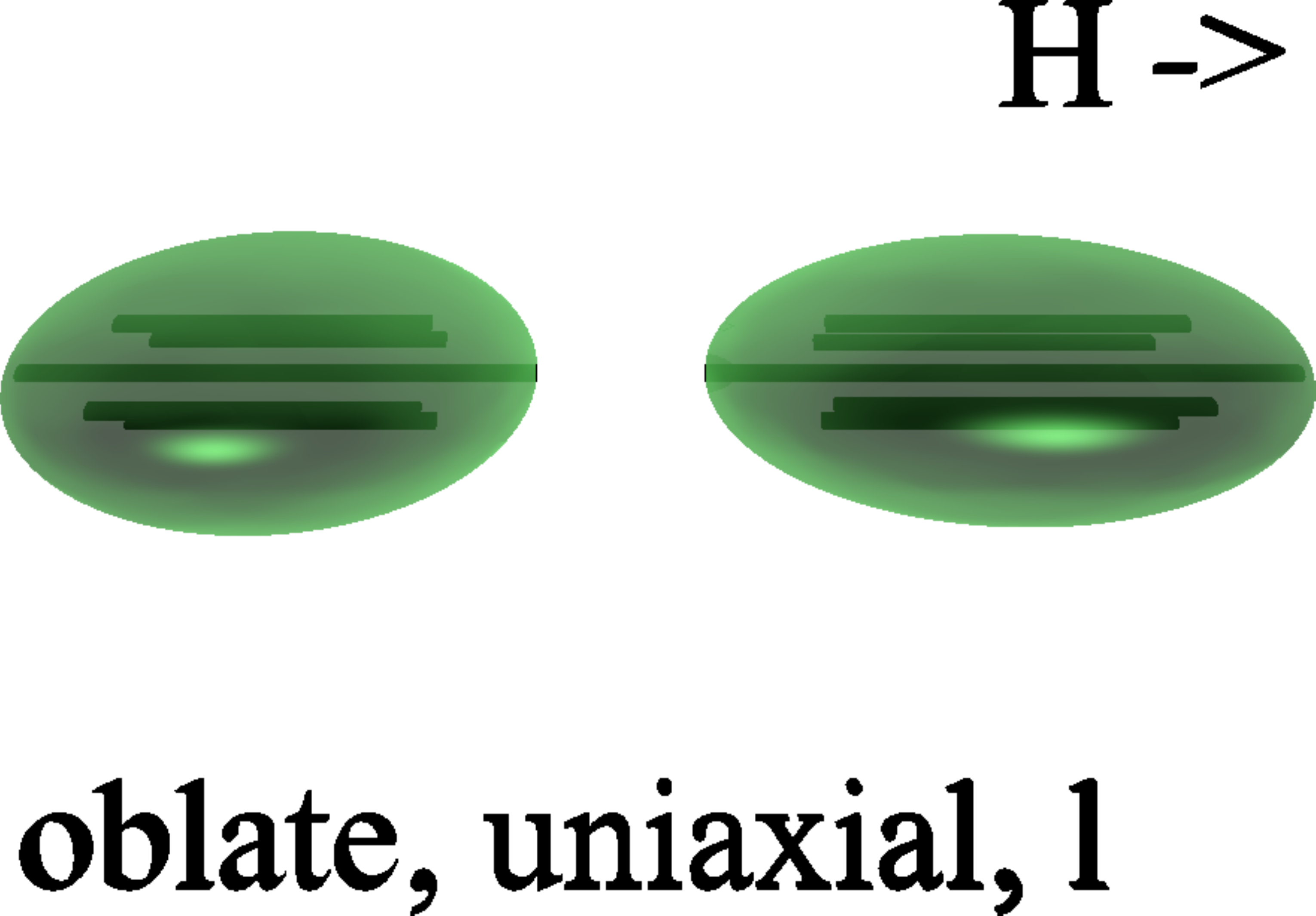}\\

\vspace{1em}
\includegraphics[width=0.26\linewidth]{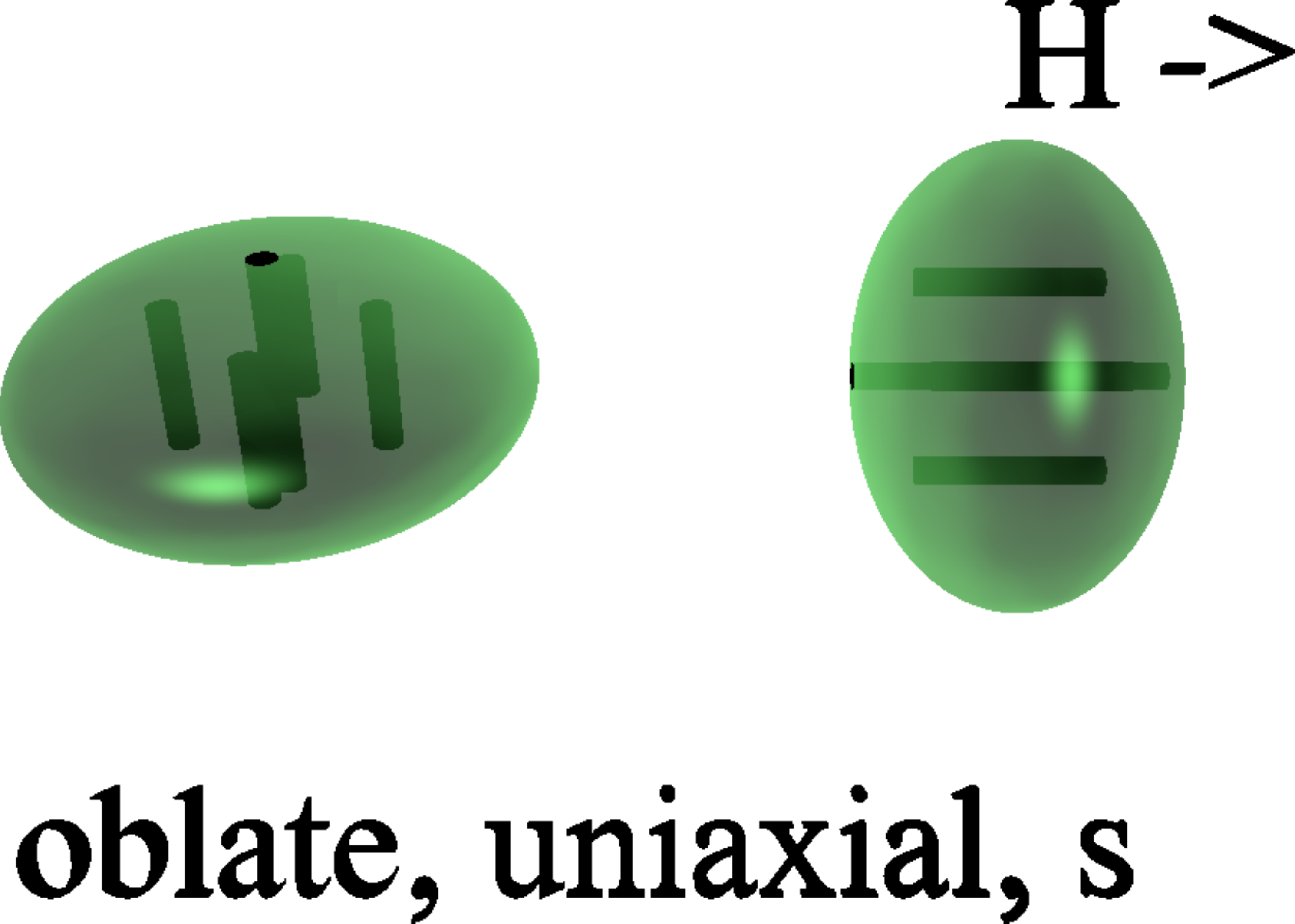}
\hspace{1em}
\includegraphics[width=0.26\linewidth]{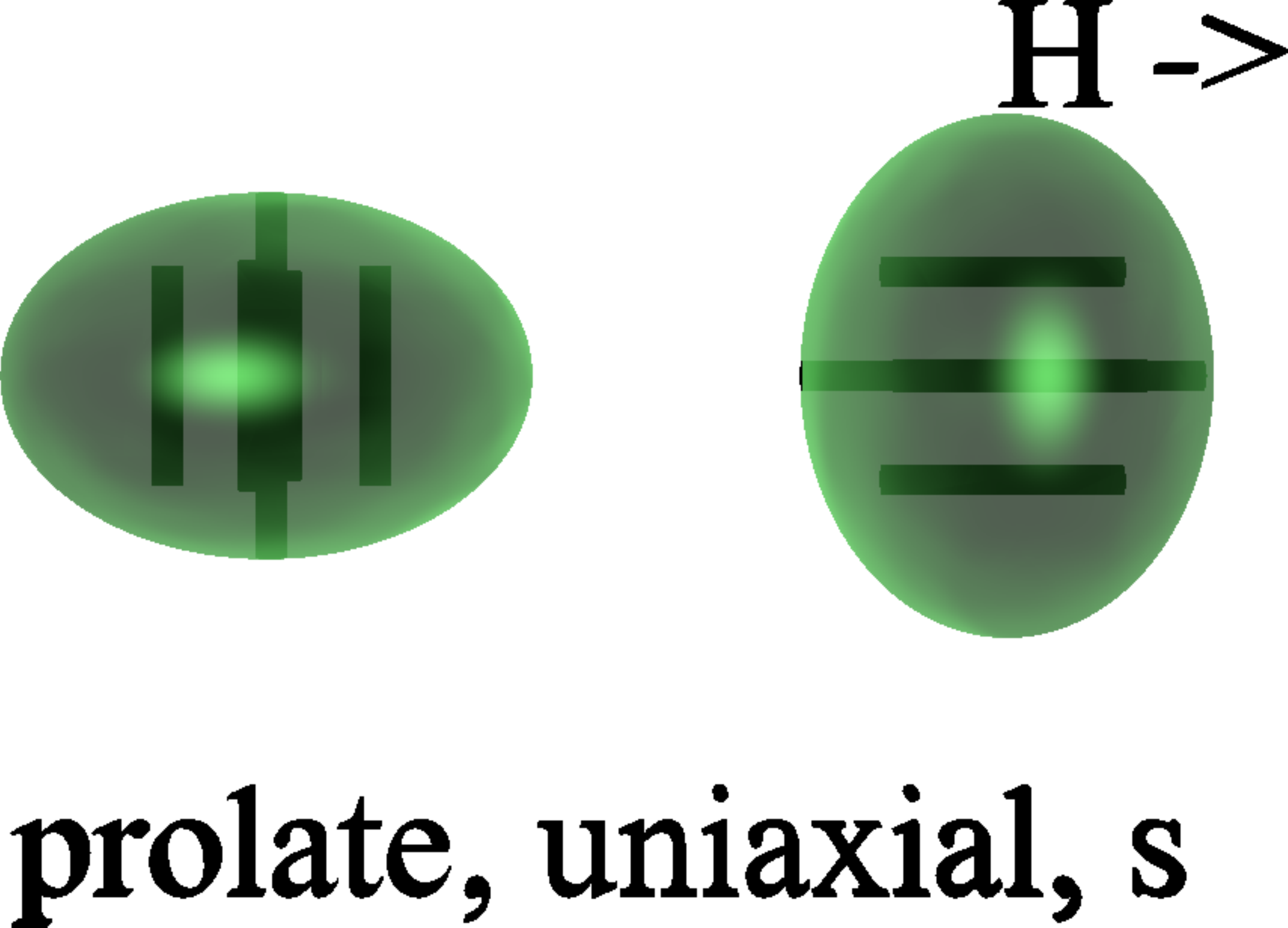}
\caption{\label{fig:alignment}
Alignment of ellipsoidal gels with isotropic and uniaxial microstructure in an external magnetic field.
Isotropic gels align one of their long axes to an external field to reduce the demagnetization energy. This is also true for uniaxial gels with a cluster structure aligned along a long axis 
If the cluster structure  of a uniaxial gel is, however, aligned along a short axis, the gel orients this axis parallel to the field. A favorable dipole-dipole and dipole-field configuration can prevent a reduction of the demagnetization energy. The deformations shown are strongly exaggerated to improve visibility.
}
\end{figure}

There are two possibilities for the ellipsoids with uniaxial microstructure, whose orientation was prescribed by an external magnetic field during cross-linking: the alignment of the microstructure can be either along the long or the short axis of the gel.
In the former case, the sample shape and the microstructure support the same orientation of a gel in a field: that with the long axis aligned parallel to the field.
When the microstructure is aligned along a short axis, however, competition arises between two different effects: the dipolar interaction between neighboring particles and the demagnetization energy can not assume their ground state simultaneously, because the former requires the microstructure (short axis) to be aligned with the field, while the latter requires the long axis to be aligned with the field.
In the next paragraphs, we  will discuss the different cases in more detail. A visual summary is presented in Fig.\,\ref{fig:alignment}.

\begin{figure*}
\includegraphics[angle=270,width=\stdfigwidth]{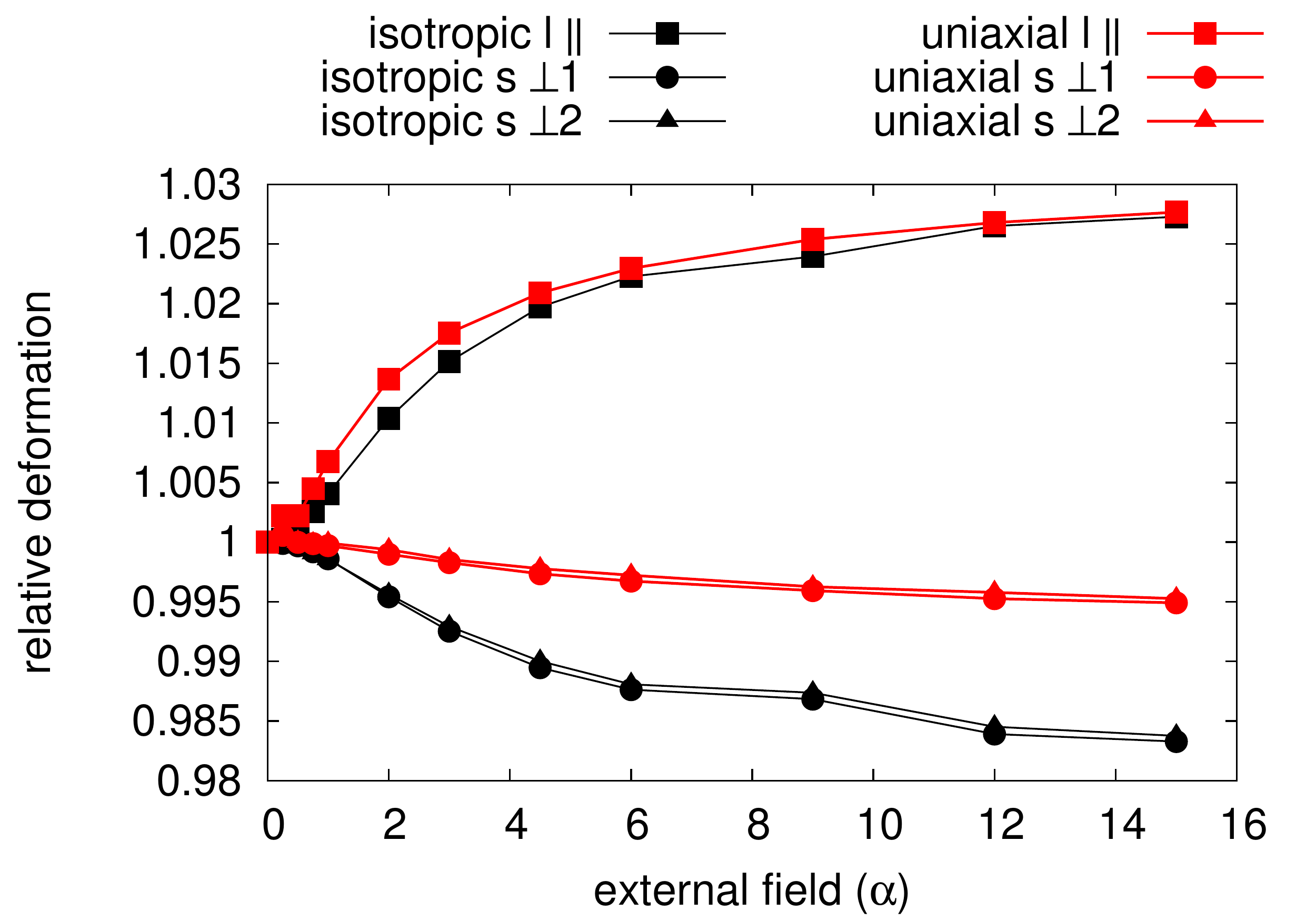}
\includegraphics[angle=270,width=\stdfigwidth]{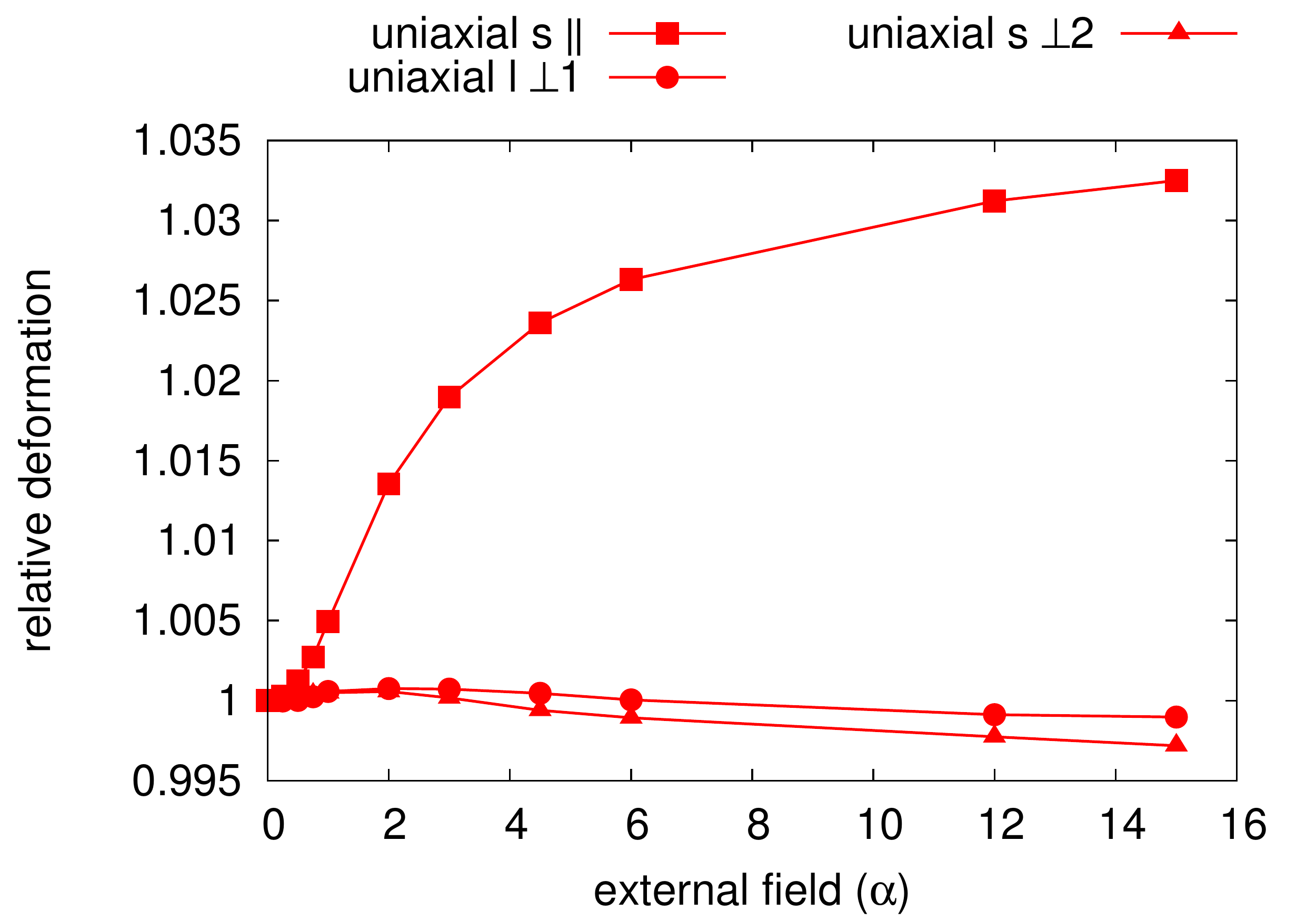}
\caption{\label{fig:prolate}
Relative deformation of prolate ellipsoidal gels. 
Here, $\parallel$ and $\perp$ denote the orientation relative to the field, l and s denote long and short axes, respectively.
The left image depicts results for an isotropic gel and a gel with a uniaxial structure parallel to the long axis. The deformation curves are qualitatively the same as for a spherical sample, as the gel always aligns its long axis parallel to the field, resulting in the same cylindrical symmetry as for a sphere.
The right hand side shows results for a sample with uniaxial structure aligned along one of the short axes. This gel aligns a short axis parallel to the field to avoid unfavorable dipole-dipole interactions between adjacent magnetic particles in the uniaxial microstructure. Hence, the cylindrical symmetry in the deformation curves is broken.
}
\end{figure*}
Let us first look at the deformation and alignment of a prolate ellipsoid. The details of the measurement procedure can be found in Appendix \ref{sec:det-shape}.
The left hand side of Fig.\,\ref{fig:prolate} shows results for an isotropic gel and a uniaxial gel with a microstructure aligned parallel to the long axis.
In the isotropic case, the gel's long axis is aligned parallel to the field due to the reduction in demagnetization energy. A uniaxial gel with a microstructure aligned along the long axis will align similarly, due to both, the demagnetization energy and the microstructure.
As a result, the system possesses cylindrical symmetry. The deformation curves are therefore qualitatively similar to those for a spherical system as shown in Fig.\,\ref{fig:spherical}: the gel elongates along the field direction. For the isotropic case, it shrinks noticeably in the perpendicular direction, while it remains largely undeformed in the uniaxial case.
On the right hand side of Fig.\,\ref{fig:prolate}, we consider a uniaxial prolate gel for which the microstructure is aligned perpendicular to the long axis. In that case, the demagnetization energy and the microstructure individually favour different alignments in the external field. For the shape considered here, the microstructure dominates and the gel aligns a short axis along the field direction. This system does not possess a cylindrical geometry, and all three axes deform to different extents.

\begin{figure*}
\includegraphics[angle=270,width=\stdfigwidth]{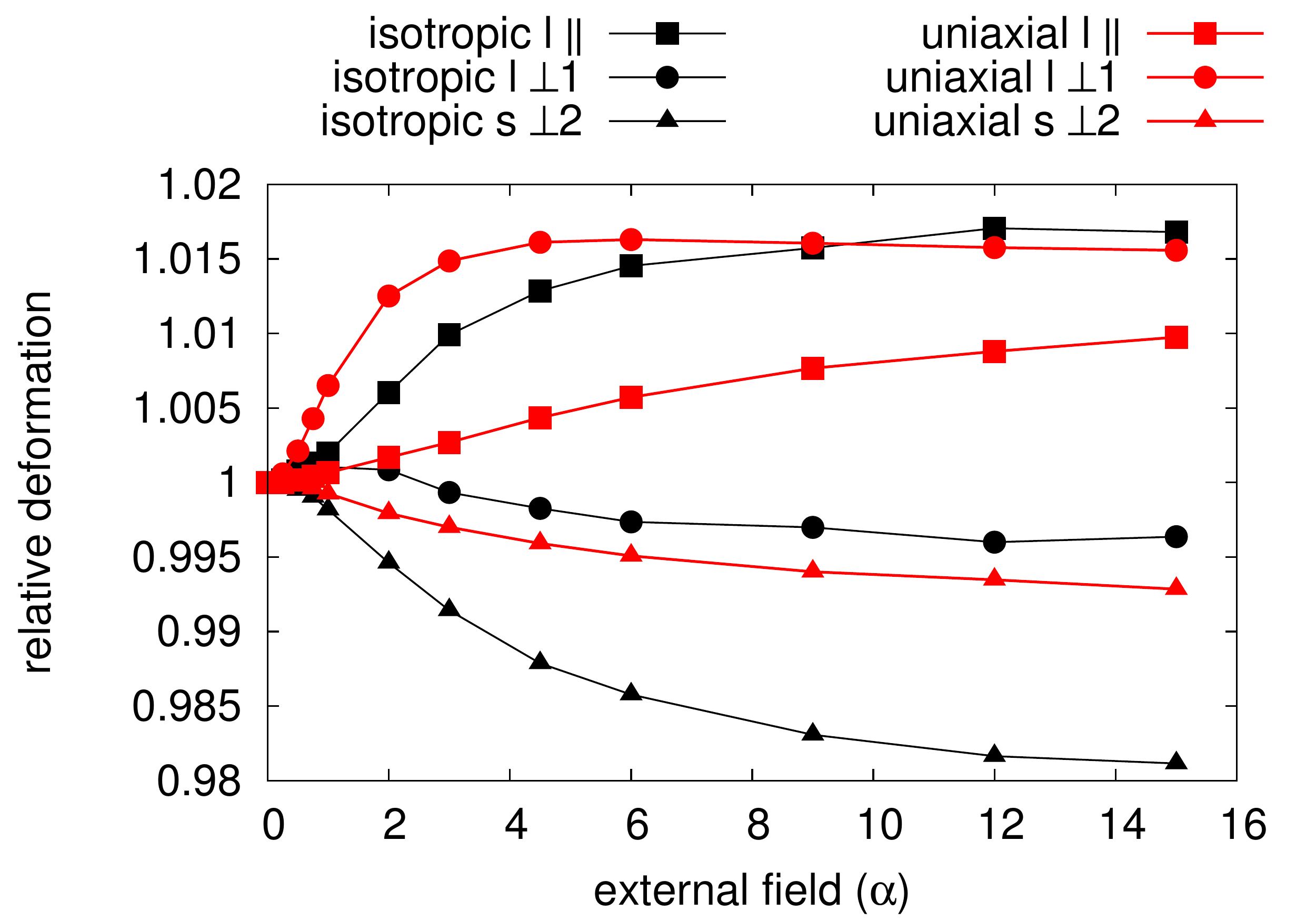}
\includegraphics[angle=270,width=\stdfigwidth]{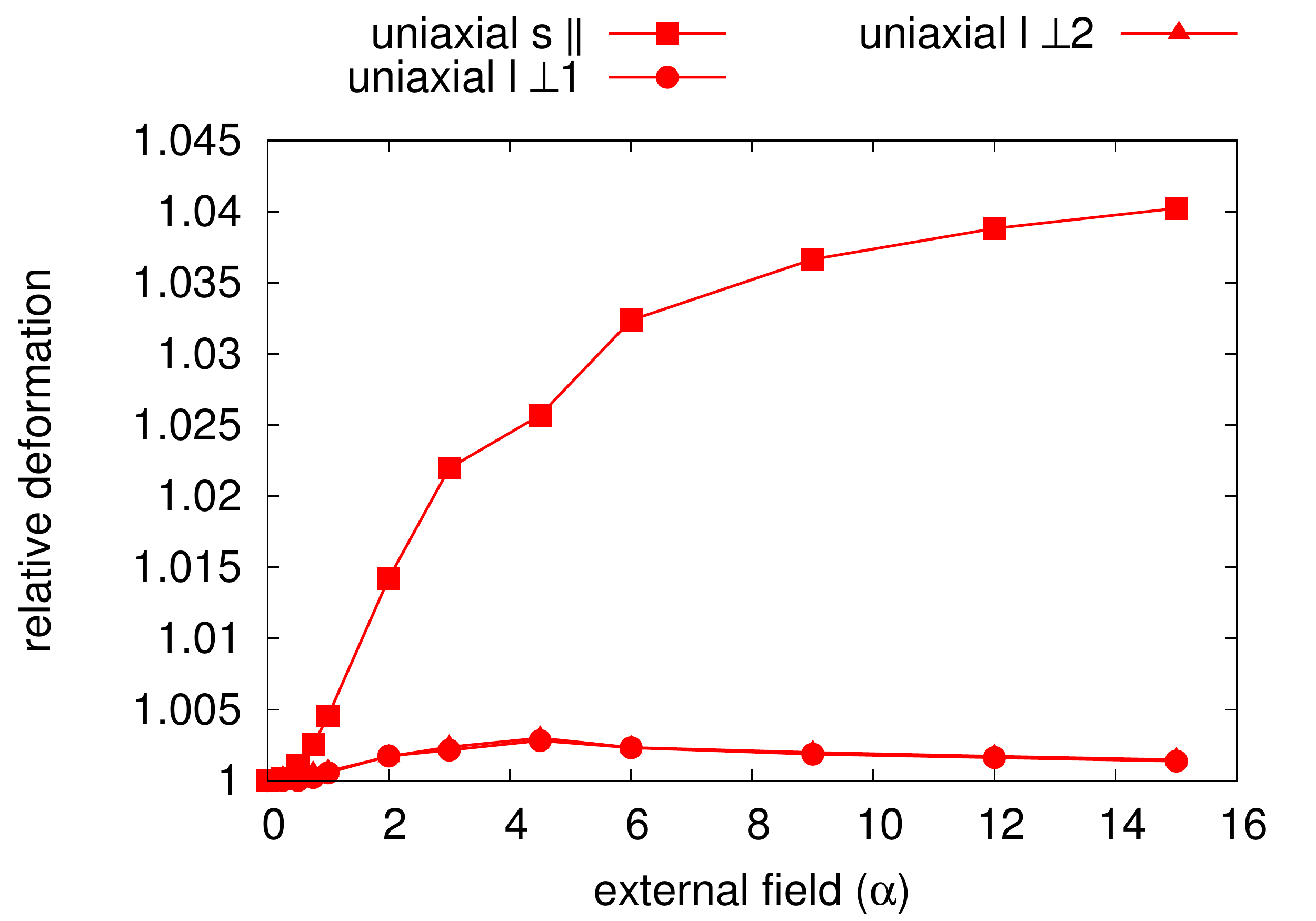}
\caption{\label{fig:oblate}
Relative deformation of oblate gel samples. 
Again, $\parallel$ and $\perp$ denote the orientation relative to the field, l and s denote long and short axes, respectively.
On the left, we show results for an isotropic gel and a gel with a uniaxial structure along the long axis. 
The gel aligns a long axis parallel to the field. There is no cylindrical symmetry, hence a different deformation for all three axes. 
On the right, we show results for a uniaxial gel with a microstructure aligned to the short axis. Here, the microstructure results in an alignment of the short axis parallel to the field, in spite of the disadvantageous demagnetization energy. The system has cylindrical symmetry, which is observed in the deformation curve, i.e., the symbols for the two perpendicular cases coincide.
}
\end{figure*}
Fig.\,\ref{fig:oblate} depicts the deformation of oblate samples.
Here, the gel's long axis aligns with the field direction if the gel is isotropic or possesses a uniaxial structure aligned along the long axis (left panel). In contrast to the prolate case, the cylindrical symmetry is broken, and hence the deformation for all three axes is differs.
On the right hand side, results for a uniaxial structure aligned along the oblate's short axis are shown. In this case, the gel aligns its short axis with he field. This system again possesses cylindrical symmetry, consequently the results are qualitatively similar to the uniaxial spherical case in Fig.\,\ref{fig:spherical}.

In summary, the microstructure of an ellipsoidal gel, if it is aligned along a short axis, can control the alignment of the gel in a field and result in a deformation behaviour different from what would result from a consideration of the demagnetization energy alone.
Fig.\,\ref{fig:alignment} summarizes the resulting alignments for the different cases.

\section{Magnetic response}
\label{sec:mag}
\begin{figure}
\includegraphics[angle=270,width=\stdfigwidth]{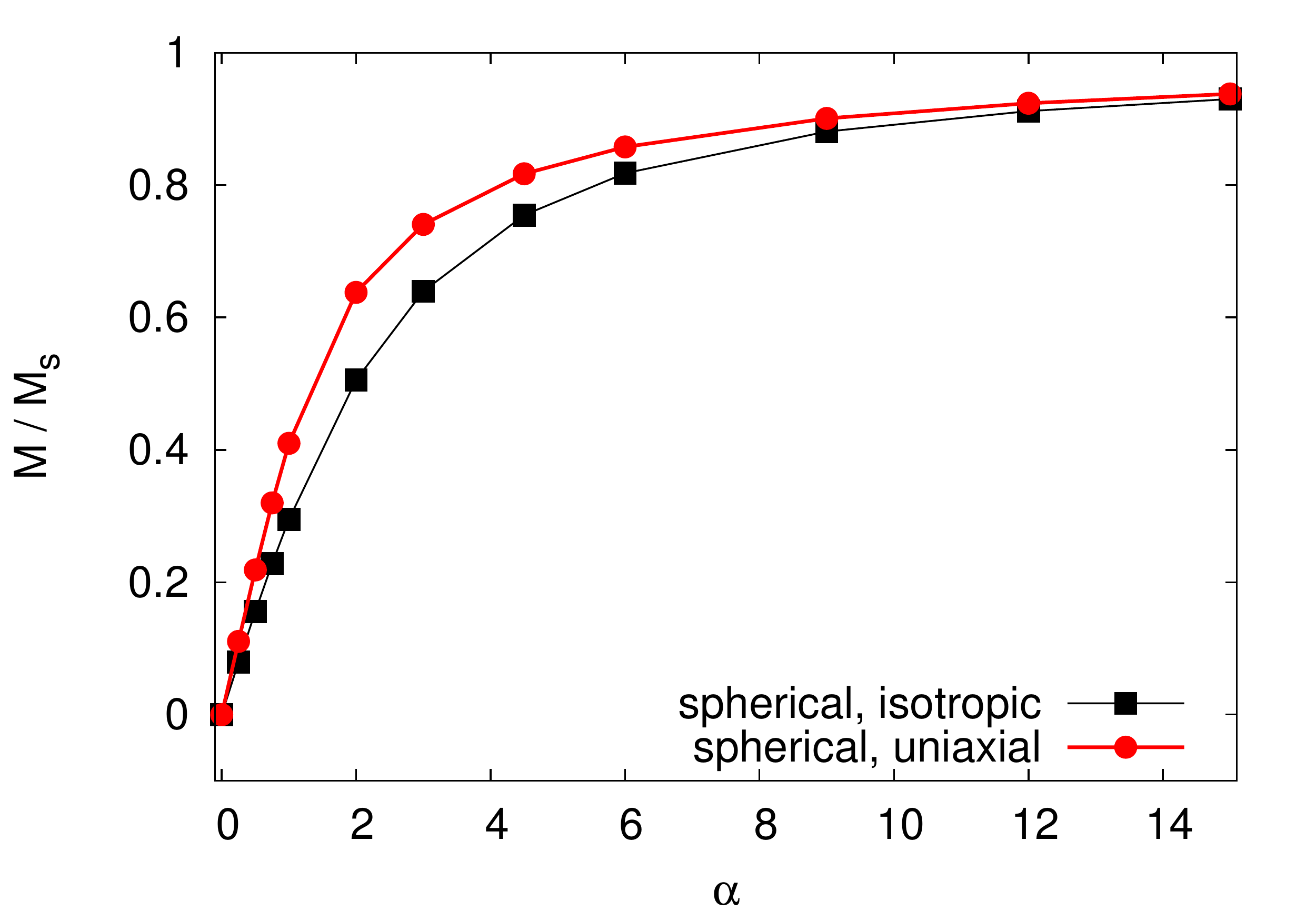}
\caption{\label{fig:mag}
Magnetization curve for a spherical gel. We observe the typical shape for a super-paramagnetic system, namely a steep initial increase and a saturation for strong external fields.
Comparing samples with isotropic and uniaxial microstructures, one find a larger initial slope for the uniaxial case, as the pre-aligned chains of magnetic particles support the magnetization.
For the more complex cases of ellipsoidal samples, the initial susceptibilities are shown in Fig.\,\ref{fig:susc}.
}
\end{figure}
\begin{figure}
\includegraphics[angle=270,width=\stdfigwidth]{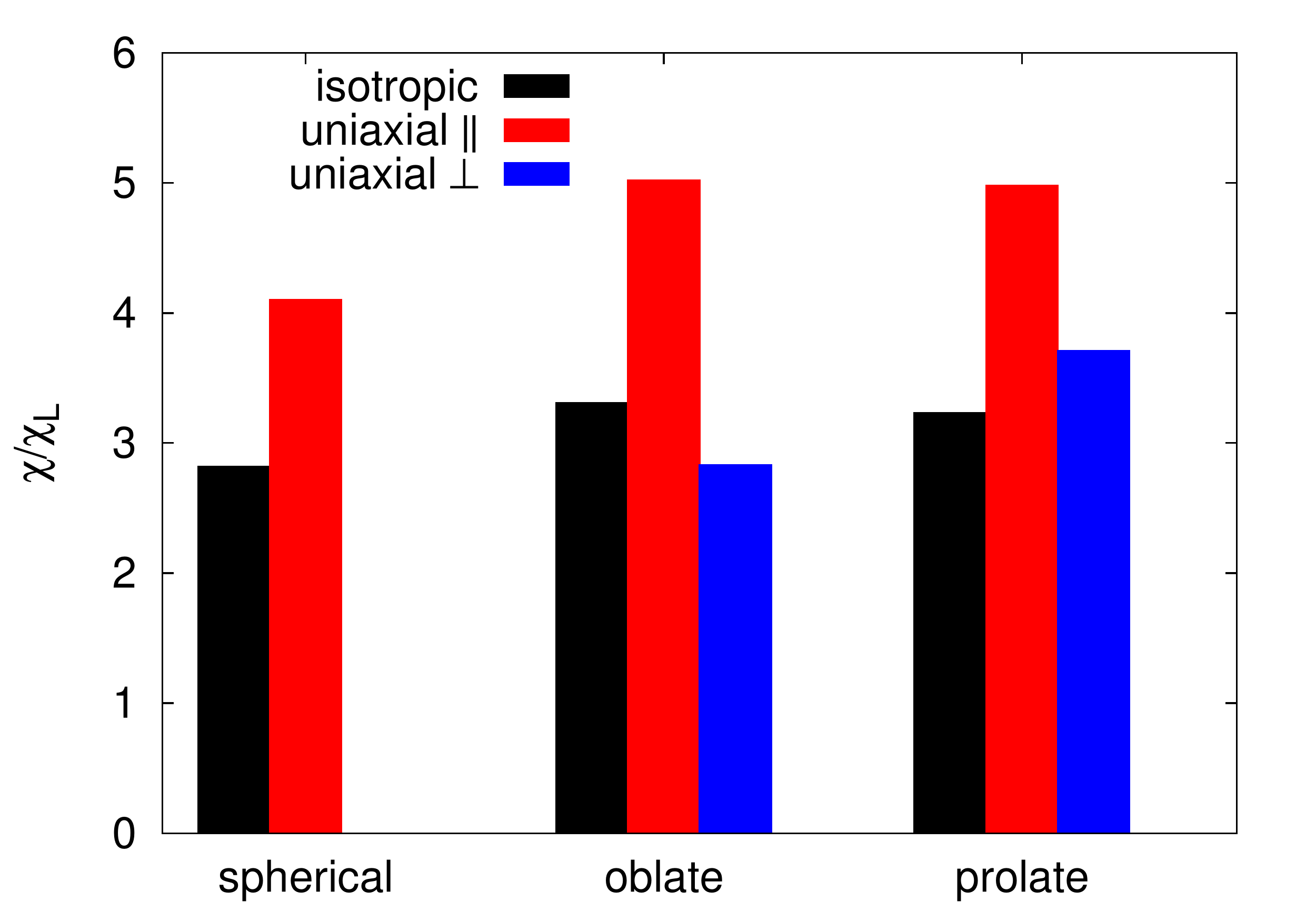}
\caption{\label{fig:susc}
Initial susceptibility $\chi$ (Eqn.\,\ref{eqn:susc}) for spherical, oblate, and prolate gels, normalized to the Langevin susceptibility $\chi_L$ for non-interacting dipoles. For each of these shapes, we compare an isotropic microstructure to uniaxial ones with a microstructure aligned parallel and perpendicular to the gels' long axes, respectively.
For all shapes, gels with a uniaxial microstructure parallel to the long axis significantly increases the susceptibility, as the pre-aligned chains of magnetic particles support the magnetization of the gel. When the microstructure has a preferred direction along a short axis of the gel, this increase of susceptibility does not occur the benefit provided by the chains of magnetic particles is compensated by the conflicting tendencies to align the gel parallel to either the microstructure direction or the gel's long axis due to the demagnetization energy.
}
\end{figure}

In this section, we show how the interplay between the gel's microstructure and its shape also influences the magnetic response, which we study using initial susceptibilities.
As before in Section \ref{sec:microstructure} and \ref{sec:shape}, we consider spherical, oblate, and prolate shapes.
Concerning the microstructure, we compare the isotropic case, with gels that have a uniaxial structure aligned parallel and perpendicular to the gel's long axis.

The magnetization curves show the sum of all dipole moments projected onto the field direction versus the strength of the external field, i.e., 
\begin{equation}
M(\alpha) =\sum_i \langle \bm{m_i} ,\bm{\hat{\alpha}} \rangle,
\end{equation}
where the $\bm{m_i}$ denote the individual particles' magnetic moments and $\bm{\hat{\alpha}}$ is the unit vector in the direction of the external field. 
We average over 3000 snapshots of the gel for a given field.
The magnetization curves of the gels have the typical shape known from super-paramagnetic systems, i.e., a steep initial ascent and a saturation for strong external fields. As an example, results for the spherical case are shown in Fig.\,\ref{fig:mag}.

For a comparison of different samples, rather than looking at magnetization curves, it is helpful to compare their initial slope, the so-called initial susceptibility.
We obtain it by fitting a Langevin-type equation
\begin{equation}
L(\alpha) =\frac{\cosh{c \alpha}}{\sinh{c \alpha}} -\frac{1}{c \alpha}
\end{equation}
to the initial part of the magnetization curve (for fields up to $\alpha=1$). The susceptibility is then given by
\begin{equation}
\label{eqn:susc}
\chi =\lim_{\alpha \to 0} \frac{\partial L(\alpha)}{\partial \alpha}=c/3.
\end{equation}
Here, $c$ is the fit parameter. It is worth mentioning that the fitted Langevin curve does not describe the entire magnetization curve, it is just used to determine the initial slope.
Results are shown in Fig.\,\ref{fig:susc}. The values shown are normalized to the Langevin susceptibility, which is the value for a system of non-interacting dipoles.

For the spherical case, we can see that the presence of a uniaxial microstructure significantly increases the susceptibility. This is the case because for a uniaxial gel, no re-alignment of individual chains of magnetic particles is needed. Rather, the gel rotates as a whole to align the chains parallel to the external field. The chains oriented parallel to the field are then very easily magnetizable, hence the high susceptibility.

Let us now turn to the ellipsoidal sample shapes. Here, we find the same increase of susceptibility for a uniaxial gel, if the gel's microstructure is aligned parallel to the gel's long axis.
If the preferred direction of the microstructure is along a short axis, however, the increase does not occur and the susceptibility can even be slightly lower than for the isotropic case.
This observation can be explained by noting the conflict between aligning the microstructure parallel to the field and thereby optimizing the dipole-dipole interaction between neighboring particles on the one hand, and orienting a long axis parallel to the field to reduce the demagnetization field, on the other hand.
Both, the dipole-dipole interaction and the demagnetization energy become the more relevant, the more the material is magnetized. Thus, the conflicting tendencies to align the microstructure and to reduce the demagnetization energy hinder the material's magnetization and lower the initial susceptibility.
In this way, the magnetization behavior supports the findings described in the previous section.

\section{Network topology}
\label{sec:topology}

As we saw in the previous sections, their can be quite strong differences between isotropic and uniaxial gels concerning their alignment and deformation in a magnetic field as well as their magnetic properties.
One open question is, however, to what degree the microstructure present during the cross-linking process is maintained after cross-linking. Thus, for instance, how ``uniaxial`` does the microstructure stay, after the field which was applied during cross-linking is removed, and how does this depend on the way the gel is cross-linked.
In this section, we first suggest a quantitative measure for how uniaxial a gel sample is, and then use it to compare different network topologies.

When quantifying how uniaxial the configuration of magnetic particles is, we cannot use the particles' dipole moment, as it is random in the absence of an external field. The initial susceptibility can also not be used, as we saw in the previous section that it depends not only on the microstructure but also on the sample shape and the relative alignment of the microstructure.
Hence, we use a measure, which is based on the cluster structure of the magnetic particles. For the cluster analysis, the same procedure is used as in Sec.\,\ref{sec:microstructure}.
Recall that clusters in magnetic fluids are often chain-like, hence, a cluster tends to have a preferred direction.
We determine this preferred direction from the clusters' inertia tensors (Eqn.\,\ref{eqn:itensor}): the rotation axis corresponding to the lowest moment of inertia is identified with the clusters ``long`` axis. 
Thus, for each cluster, we obtain an orientation $\bm{\hat{o}_c}$, by taking the normalized eigenvector corresponding to the lowest eigenvalue of the inertia tensor.
For a given direction $\bm{\hat{x}}$, the nematic order parameter is
\begin{equation}
\phi(\mathbf{\hat{x}}) =\frac{1}{N_c} \sum_c \frac12 \left(3 \langle \mathbf{\hat{o}_c}, \mathbf{\hat{x}} \rangle^2 -1 \right),
\end{equation}
where $N_c$ denotes the number of clusters, and $\bf{\hat{o}_c}$ the normalized orientation vector of each cluster, respectively.
The order parameter, which stems from the study of liquid crystals\cite{eppenga84a}, takes a value of $1$, when all clusters are aligned parallel or anti-parallel to $\bf{\hat{x}}$, $0$ when they are randomly aligned and $-0.5$, when aligned perpendicular.
We then define the degree of uniaxiallity $d_u$ of the microstructure as
\begin{equation}
\label{eqn:uniaxiallity}
d_u =\rm{max}_{\bf{\hat{x}}} \phi(\bf{\hat{x}}),
\end{equation}
i.e., as the maximum nematic order parameter of the cluster orientation vectors for all given directions (Recall that the orientation of the gel in the simulation box is not known a priori, if no field is applied).

\begin{figure}
\includegraphics[angle=270,width=\stdfigwidth]{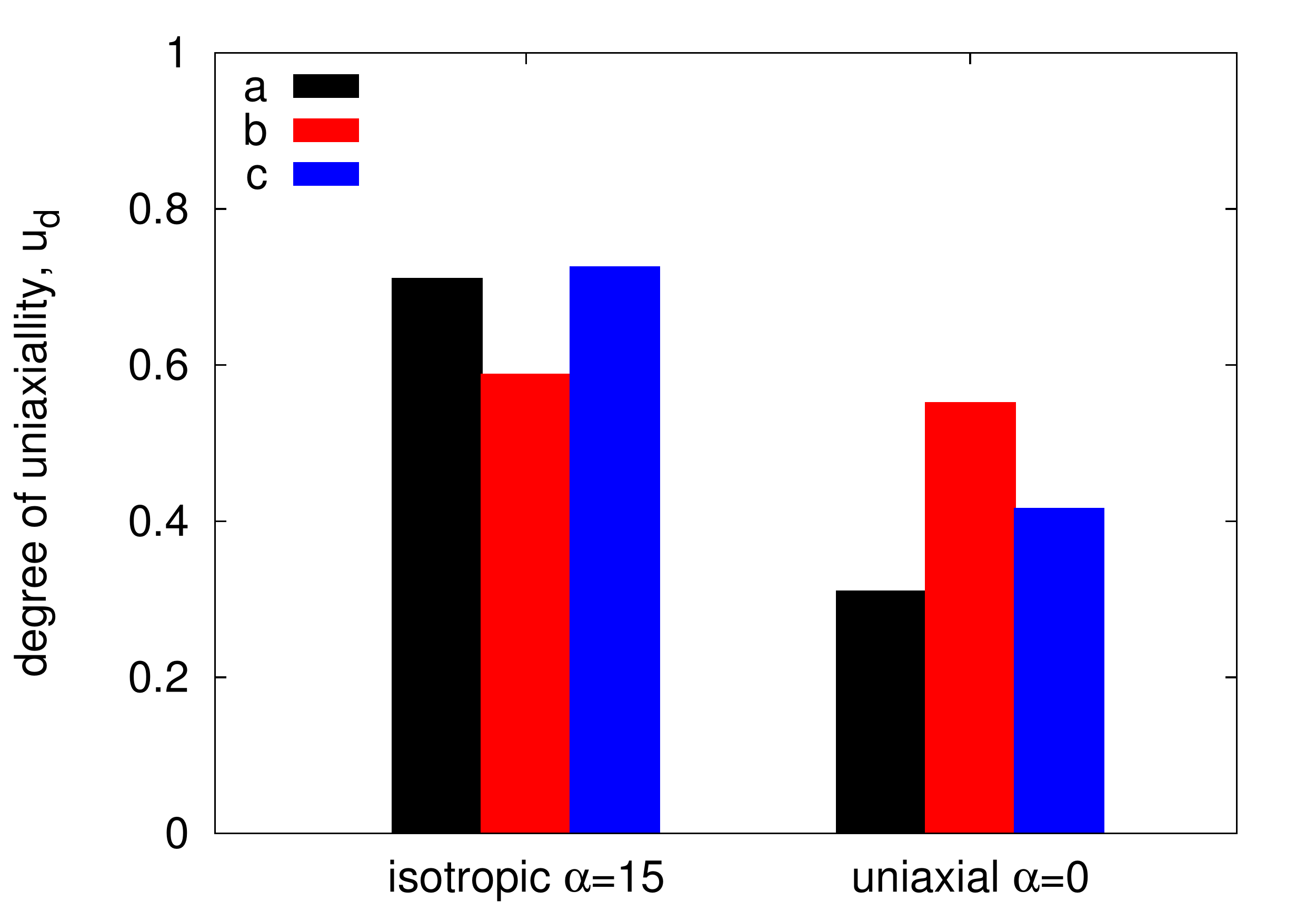}
\caption{\label{fig:uniaxiallity}
Degree of uniaxiallity, $d_u$, Eqn.\,\ref{eqn:uniaxiallity}, for an isotropic gel placed into a field and for a uniaxial gel observed in the absence of a field.
We compare results for three network topologies (Tbl.\,\ref{tbl:bond-parameters}, Fig.\,\ref{fig:bond-length-dist}).
In both cases, networks in which mainly neighboring particles are bound (case b), retain the initial structure most.
This implies a lower degree of uniaxiallity for an isotropic gel in a field and a higher one for an isotropic gel in the absence of a field.
}
\end{figure}

Let us now use this measure to examine gels with different network topologies as defined in Eqns.\,\ref{eqn:bond-prob}-\ref{eqn:bond-stiffness} and Tbl.\,\ref{tbl:bond-parameters}.
We compare three cases, $a$ through $c$. Cases $a$ and $c$ refer to a gel where particles several diameters apart are the most likely to be connected by a polymer. Here, case $a$ uses a narrow distribution of bond lengths and case $c$ a wide one. Case $b$, on the other hand, pertains to a situation where polymer connect mainly closely neighboring particles.
In Fig.\,\ref{fig:bond-length-dist}, the bond length distributions (Eqn.\,\ref{eqn:bond-length-dist}) for the three cases are shown.
In Fig.\,\ref{fig:uniaxiallity}, we show the corresponding degrees of uniaxiallity $d_u$, (Eqn.\,\ref{eqn:uniaxiallity}) for two situations, namely an isotropic gel to which an external field is applied and a uniaxial gel, to which no field is applied after cross-linking.
If a gel has a flexible microstructure, this implies a high degree of uniaxiallity for an isotropic gel, to which a field is applied. Conversely, the uniaxiallity of a uniaxial gel is low, when the field is released, if there is a lot of flexibility for the particles to rearrange.
For the three network topologies compared in Fig.\,\ref{fig:uniaxiallity}, we find that for the cases $a$ and $c$, the particles have more flexibility to rearrange than for the case $b$.
From this observation one may draw the conclusion that binding magnetic particles close to each other tends to stabilize the network structure more than polymers connecting more distant particles.

\section{Summary}

In this article, we proposed a model for a ferrogel suitable to study effects related to the particle microstructure, sample shape, and polymer network topology.
Concerning the particle microstructure, we showed that in an isotropic gel, the application of an external field reduces clustering as, in such a gel, chains of magnetic particles are not able to align to the field direction. The interactions between particles in such chains become repulsive, if the magnetic moments are aligned with the field.
When non-spherical gels are considered, the demagnetization energy is of importance. In order to reduce it, a long axis of the gel is aligned parallel to the field, if the sample has an isotropic microstructure. For a sample with a uniaxial microstructure, however, a conflict can arise between aligning the chains of magnetic particles parallel to the field, on the one hand, and aligning a long axis parallel to the field, on the other hand. For the oblate and prolate ellipsoid considered in this report, the orientation of the microstructure determines the final alignment of the gel.
This is plausible as not aligning the preferred direction of the microstructure with the field would result in a large number of adjacent particles with energetically very unfavorable interactions. Not aligning a long axis, on the other hand means that particles with unfavorable alignments are far apart.
Finally, we turned to the influence of the polymer network topology. To study its effect, we introduce a measure to quantify the degree to which a gel has a uniaxial particle microstructure. With this measure, we observe the influence of the network topology on the stability of the microstructure. Our findings suggest that the microstructure is more stable, if the polymers link closely located, adjacent magnetic particles. This may be of interest for applications, which are based on a uniaxial particle microstructure. In that case, it is important to ensure that the microstructure of the gel maintains the uniaxiallity, when the external field is removed after cross-linking.
In summary, we have suggested a model for a magnetic gel, which is suitable to describe situations, where the particle microstructure, the sample shape, and the network topology are of similar importance.

\begin{appendix}
\section{Technical description of the simulations}
\label{sec:sim}

As explained in Sec.\,\ref{sec:model}, the simulations consist of four steps.
First, the microstructure of a ferrofluid is obtained, then the sample shape is cut, the gel is cross-linked and finally observables for the cross-linked gel are recorded.

\subsection{Obtaining the microstructure of a ferrofluid}
\label{sec:sim-ff}

We perform molecular dynamics simulations by means of ESPResSo\cite{limbach06a,arnold13a}.
In a system with a volume fraction of 5\%, we simulate 20\,000 magnetic particles with a dipolar interaction parameter (Eqn.\,\ref{eqn:lambda}) of $\lambda=4$.
To obtain the microstructure of a uniaxial gel, we additionally apply a magnetic field of $\alpha=15$, where $\alpha$ is the Langevin parameter given by
\begin{equation}
\alpha = \frac{\mu_0 m H}{k_B T}.
\end{equation}
The particles are modelled as soft spheres interacting via the WCA-Potential\cite{weeks71a}, which is given by
To obtain the microstructure of a uniaxial gel, we additionally apply a magnetic field of $\alpha=15$, where $\alpha$ is the Langevin parameter given by
\begin{equation}
\alpha = \frac{\mu_0 m H}{k_B T}.
\end{equation}
The particles are modelled as soft spheres interacting via the WCA-Potential\cite{weeks71a}, which is given by
\begin{equation}
\label{eqn:wca}
U_{\textsc{wca}} = 
\begin{cases}
4\epsilon
\left[\left(\frac{\sigma}{r_{ij}}\right)^{12}
 - \left(\frac{\sigma}{r_{ij}}\right)^{6}\right]+\epsilon & r<r_c\\
0 & r \geq r_c
\end{cases}
\end{equation}
We use $\sigma=1$ as particle diameter and $\epsilon=10$ as energy scale.
We employ the $NVT$ ensemble, and both, translational and rotational degrees of freedom are thermalized by means of a Langevin thermostat with a thermal energy of $k_B T=1$ and a friction coefficient $\gamma=1$. 
Periodic boundary conditions are applied. The dipole-dipole interaction is calculated using the P3M method\cite{cerda08d}, tuned to a mean absolute force error of $10^{-4} k_B T / \sigma$. The system is integrated for 100\,000 time steps of size $dt=0.003$, before the particle structure is captured.

\subsection{Cutting a shape and cross-linking}

From the particle configuration obtained in the previous step, the desired shapes are cut. In this paper, we study spheres as well as prolate and oblate ellipsoids of revolution.
After the cutting, the gel is cross-linked by randomly adding harmonic bonds between the particles based on the probability function given in Eqn.\,\ref{eqn:bond-prob}, the bond stiffness given in Eqn.\,\ref{eqn:bond-stiffness} and the parameters from table \ref{tbl:bond-parameters}.
The bonds are described by a harmonic potential
\begin{equation}
U(r) =\frac12 k(r_0) \langle{ r-r_0 \rangle}^2.
\end{equation}
The long axes of all ellipsoids considered is equal to the diameter of the spherical system $d$. Oblate and prolate ellipsoids contain approximately half the number of particles of the spherical system. The short axis of the oblate system is $0.5d$, those of the prolate system are $\sqrt{0.5}d$ in length.

\subsection{Computing observables for the cross-linked gel}
\label{sec:det-shape}

The simulations of the final gel are based on those described in appendix \ref{sec:sim-ff}.
There are, however, two crucial differences.
First, the simulations are carried out in open rather than in periodic boundary conditions in order to capture the sample shape. This implies that the P3M method can no longer be used to calculate the dipolar interactions. Instead, we use direct summation. Due to the large number of approximately 100~million pairs to be considered, the calculations are run on a graphics processor.
Second, the molecular dynamics simulations discussed above are supplemented by Monte Carlo stretching and rotation moves to provide a faster traversal of configuration space.
Alternating, every 100 time steps of size $dt=0.01$, a stretching or a rotation move is performed.
In a stretching move, each Cartesian coordinate is altered by $\pm \delta$, where $\delta$ is drawn from a uniform random distribution in the interval of $-0.005\sigma \to 0.005\sigma$.
In a rotation move, the system is rotated around a randomly picked Cartesian axis by an amount $\delta_r$ drawn from a uniform distribution in the interval $-0.2\to0.2$.
The rotation comprises both, particle positions and dipole moments.
The moves are accepted or rejected based on the metropolis criterion:
if the change in energy due to the move is less than zero, the move is accepted. If it is larger than zero, the move is accepted if $R<=\exp{-\delta E / (k_B T)}$, where R is a random number drawn from a uniform distribution in the interval $0\to1$.
It is not trivial to determine the ``shape`` of an assembly of point particles.
In cases, where the location of the interface needs to be known with a high accuracy, a hull can be constructed\cite{sega13d}.
For the purpose of obtaining the deformation of our model gels, only knowledge about the change in elongation is needed, allowing us to use a simpler approach.
As the spherical and ellipsoidal gels deform by less than five percent, we can approximate their shape as ellipsoids, even after a deformation has taken place.
We obtain the three axes defining the ellipsoid by comparing the inertia tensor of the gel to that of an ellipsoid. 
This can be done using the eigenvectors of the inertia tensor. The three axes are then given by
\begin{align}
\label{eqn:shape}
l_1 = \frac{\sqrt{-10  (E_1 -E_2 -E_3)}}{2}, \\
l_2 = \frac{\sqrt{-10  (E_2 -E_1 -E_3)}}{2}, \\
l_3 = \frac{\sqrt{-10  (E_3 -E_1 -E_2)}}{2},
\end{align}
where
$E_i$ denote the eigenvalues of the inertia tensor
\begin{equation}
\label{eqn:itensor}
J =m \sum_k \left( \langle \vec{r}_k , \vec{r_k} \langle I  -\vec{r}_k \otimes \vec{r}_k \right),
\end{equation}
where $m$ is the particle mass and $\vec{r}_k$, denotes the position of the $k$'th particle. Lastly, $I$ is the identity matrix.
The orientation of the ellipsoid can be determined from the eigenvectors. 
The eigenvalue-based method is used for all ellipsoidal samples, considered.

For a spherical system without an external field, this method is not optimal, because the eigenvectors are degenerate and are only fixed by small random anisotropies in the system.
It is then preferable to work with the inertial moments around the fixed Cartesian axes. Due to the random rotations of the system during the course of the simulation, good averaging is achieved.
For this approach, the semi-axes of the ellipsoid are also obtained from Eqn.\,\ref{eqn:shape}, but the eigenvalues of the inertia tensor are replaced by the inertia moments for the Cartesian axes.

The measurements of the sample shapes in Figs.\,\ref{fig:spherical}, \ref{fig:prolate}, and \ref{fig:oblate} are averaged over 3000 measurements separated by 1000 molecular dynamics time steps and rive volume and stretching Monte Carlo moves.
The gel is allowed to adapt to the external field for two million time steps, before measurements begin.

\end{appendix}

\section*{Acknowledgements}
The authors thank Georg Rempfer for helpful comments.
The authors are grateful for financial support from the DFG through the
SPP 1681, and to the HLRS and BW-Unicluster for computing resources.

\bibliographystyle{unsrt}
\bibliography{icp}

\end{document}